\documentclass[aps, prl, reprint, a4paper, amsfonts, amssymb, amsmath,  showkeys, nofootinbib, twoside, superscriptaddress,longbibliography]{revtex4-2}
\usepackage[english]{babel}
\usepackage[utf8]{inputenc}
\usepackage{cancel,amsmath}
\usepackage{amsthm}
\usepackage{mathtools}
\usepackage{xcolor}
\usepackage{graphicx} %
\usepackage{float}
\usepackage{soul}
\usepackage{gensymb}
\usepackage{multirow}
\usepackage{adjustbox}
\usepackage{placeins}
\usepackage[T1]{fontenc}

\usepackage{csquotes}

\usepackage[pdftex, pdftitle={Imaginary-Time Density Correlations Encode Quantum Entropy and Free Energy}, pdfauthor={Sagi Meir}]{hyperref} %
\usepackage[capitalize]{cleveref}

\crefname{section}{Sec.}{Secs.}
\Crefname{section}{Section}{Sections}
\crefname{smsection}{Sec.}{Secs.} %
\Crefname{smsection}{Section}{Sections}
\begin{document}
\title{Imaginary-Time Density Correlations Encode Quantum Entropy and Free Energy}
\author{Sagi Meir}
    \email[Author to whom any correspondence should be addressed: ]{sagimeir@tauex.tau.ac.il}
    \affiliation{School of Chemistry, Tel Aviv University, Tel Aviv 6997801, Israel}
    \affiliation{The Center for Physics and Chemistry of Living Systems, Tel Aviv University, Tel Aviv 6997801, Israel}

\begin{abstract}
    In the classical limit, maximum-entropy inference yields an estimate of the excess entropy from the static structure factor through a single momentum sum. Quantum mechanically this estimate fails, because interactions also change the kinetic energy, while delocalization and exchange add information that equal-time correlations do not contain. We show that imaginary-time density correlations contain this information, and that with them the classical relation carries over to quantum systems. Such correlations are available from path-integral simulations and X-ray Thomson scattering. Relative-entropy inference over imaginary-time paths, with an ideal Bose or Fermi gas as reference, gives closed forms for the excess entropy and free energies, within the same Gaussian approximation as the classical relation. They recover the classical result and estimate entropy that the classical relation misses. We test them on weakly interacting bosons, interacting fermions on a ring, and the warm dense electron gas. The closed forms need only imaginary-time density correlations and their ideal-gas reference at a single state point, without using the interaction potential. Our approach can extend other classical entropy estimators to quantum systems.
\end{abstract}
\maketitle

\addtocontents{toc}{\protect\setcounter{tocdepth}{0}} %

A direct calculation of entropy requires the full probability distribution over microstates, or the complete density matrix in the quantum case. Such information is generally inaccessible in experiments~\cite{diamant2026measuring,elben2023randomized} and prohibitively costly to obtain in simulations~\cite{frenkel2023understanding,feynman1982simulating}. Entropy is therefore usually inferred indirectly~\cite{pohorille2010good}, e.g., through thermodynamic integration along a path that connects the system to a non-interacting reference state~\cite{kirkwood1935statistical,frenkel1984new}. Such calculations typically require energies or correlation functions at every coupling between zero and the physical value, as in the coupling-constant integration for a classical fluid~\cite{hansen2006theory} or the adiabatic connection for an electron gas~\cite{groth2017ab}. For electrons, path-integral Monte Carlo simulations along this path are limited by the fermion sign problem~\cite{troyer2005computational,chuna2025second}.

Recent works developed alternative routes to estimate the entropy of classical systems through its connection to information. Instead of trying to reconstruct the full distribution of microstates, they quantify how much thermodynamic information is encoded in the spatial arrangement of particles. Examples of such entropy estimators include those based on compression algorithms~\cite{martiniani2019quantifying,avinery2019universal}, neural-network estimates of mutual information~\cite{nir2020machine,benshimon2026machine}, and pair correlations~\cite{ariel2020inferring,sorkin2023resolving,sorkin2023detecting}. At the pair level, this information is captured by the static structure factor, $S(\mathbf q) = \frac{1}{N} \left\langle \rho_{\mathbf q}\rho_{-\mathbf q} \right\rangle$, constructed from the Fourier component of the particle density, $\rho_{\mathbf q} = \sum_{j=1}^{N} e^{-i\mathbf q\cdot\mathbf r_j}$, where $N$ is the number of particles, $\mathbf r_j$ is the position of particle $j$, and $\langle\cdot\rangle$ denotes an ensemble average. Relations such as the pair-correlation expansion of the excess entropy~\cite{green1952molecular,nettleton1958expression}, excess-entropy scaling~\cite{rosenfeld1977,dyre2018perspective}, and the relation between entropy and kinetics~\cite{sorkin2023universal} connect this spatial structure to thermodynamic and transport properties.

\citet{ariel2020inferring} formulated this connection as a maximum-entropy inference problem, applicable in and out of equilibrium. Using a given $S(\mathbf{q})$ as a constraint, they obtained an entropy estimate $H$, which bounds the system's entropy from above and, within a Gaussian approximation, reads
\begin{equation}
    H-H_0 \approx \frac{1}{2}\sum_{\mathbf{q} \neq 0}\left(\ln S(\mathbf{q}) - S(\mathbf{q}) +1\right),
\label{eq:classic_entropy_estimation}
\end{equation}
where $H_0$ is the entropy of the ideal-gas reference, whose structure factor is $S_0(\mathbf q)=1$.

As we will show, extending this construction to quantum systems requires additional information. In the classical limit, the excess entropy can be related to spatial correlations, because the kinetic and potential energies decouple. In quantum systems, the kinetic and potential operators generally do not commute~\cite{wigner1932quantum,kirkwood1933quantum}. Interactions can therefore modify the kinetic energy as well as the spatial distribution, and the kinetic energy no longer follows from equipartition~\cite{ceperley1995path,ceriotti2016nuclear}. A simple example is a particle in a box: classically its mean kinetic energy is independent of the width, whereas quantum mechanically it grows as the box narrows, even at fixed temperature. Quantum delocalization and exchange statistics add thermodynamic information~\cite{feynman1953atomic,ceperley1995path} that is not contained in equal-time density correlations, such as the static structure factor.

We argue that the additional information is contained in density correlations along imaginary time. We find it by mapping a quantum system onto an isomorphic fictitious classical system, using the path-integral representation~\cite{feynman1965quantum,chandler1981exploiting,tuckerman2023statistical}. In this classical system, the position of each particle is replaced by a path $\left\{{\mathbf r}_j^{(\tau)}\right\}$, a curve in space parametrized by the imaginary time $0\leq\tau\leq\beta\hbar$, where $\beta$ is the inverse temperature. The spatial extent of this path describes quantum delocalization~\cite{parrinello1984study,gillan1987quantum,markland2018nuclear}. For this system of paths, the analog of the structure factor is the density-density correlation function~\cite{mahan2013many},
\begin{equation}
    S(\mathbf{q},\omega_n)=\frac{1}{N\hbar}\int_0^{\beta\hbar}\mathrm{d}\tau\, e^{i\omega_n\tau}\left\langle\rho_{\mathbf{q}}^{(\tau)}\rho_{-\mathbf{q}}^{(0)}\right\rangle ,
\label{eq:Matsubara_density-density}
\end{equation}
where $\omega_n=\frac{2\pi n}{\beta\hbar}$ is the Matsubara frequency, $n\in \mathbb{Z}$, and $\rho_{\mathbf q}^{(\tau)}=\sum_{j=1}^{N} e^{-i\mathbf q\cdot\mathbf r_j^{(\tau)}}$ is the density mode at imaginary time $\tau$. Summing over frequencies returns the equal-time quantity, $\beta^{-1}\sum_n S(\mathbf q,\omega_n)=S(\mathbf q)$, whereas the complete set $\{S(\mathbf q,\omega_n)\}_n$ resolves how the density correlations are distributed along imaginary time. The function $S(\mathbf{q},\omega_n)$ is available from path-integral Monte Carlo or path-integral molecular dynamics simulations~\cite{dornheim2023physical, tolias2024fourier}, and, in warm dense matter experiments, can be extracted from X-ray Thomson scattering~\cite{glenzer2009x, sheffield2010plasma, dornheim2022accurate, dornheim2024xray}.

Here, we show that the classical relation of \citet{ariel2020inferring}, \cref{eq:classic_entropy_estimation}, carries over to quantum systems, by using the complete set $\{S(\mathbf q,\omega_n)\}_n$. Our relation requires only $S(\mathbf q,\omega_n)$ and a non-interacting reference state, without requiring the interaction potential. The excess entropy and free energies then follow in closed form from correlations at a single state point, i.e., without an integration over the coupling constant. We benchmark the relation against weakly interacting bosons within the Bogoliubov approximation, a spinless-fermion ring, and the warm dense electron gas. Importantly, when $S(\mathbf q,\omega_n)$ is not available, we can infer it from $S(\mathbf q)$, under the mild assumption of an instantaneous interaction.

\textit{Derivation outline.}---To derive the quantum equivalent of \cref{eq:classic_entropy_estimation}, we look for a density matrix $\hat{P}$ that yields the given $S(\mathbf q,\omega_n)$ and a maximum value of the negative quantum relative entropy~\cite{vedral2002role},
\begin{equation}
    H_{\text{rel}}\equiv - D(\hat P\|\hat P_0) = -\text{tr}\left(\hat{P}\ln\hat{P}-\hat{P}\ln\hat{P}_0\right),
\label{eq:H_rel_def}
\end{equation}
with respect to a non-interacting reference state $\hat{P}_0$. We use Lagrange multipliers $\lambda_{\mathbf q,n}$ to enforce the given $S(\mathbf q,\omega_n)$ at each $\mathbf q\neq0$ and $n$. Then, we extremize $H_{\text{rel}}$ subject to these constraints. Using $\hat{P}_0$ as the prior builds the ideal-gas kinetic term and quantum statistics into the inferred state $\hat{P}$. For example, $\hat{P}_0$ can be an ideal Bose or Fermi gas with the corresponding density-density correlation function $S_0(\mathbf q,\omega_n)$ (Supplemental Material, \cref{sec:reference}).

The relative entropy follows from an effective partition function, in which the constraint appears as a term quadratic in the density (End Matter, \cref{sec:variational}). Already at this stage, a numerical solution for the multipliers gives $H_{\text{rel}}$ without further approximation. But, as in the classical derivation~\cite{ariel2020inferring}, to obtain a closed-form relation, we treat the density fluctuations as Gaussian; the neglected terms are three- and higher-point density correlations. This Gaussian approximation, which is the same as in \cref{eq:classic_entropy_estimation}, is the sole approximation of the framework.

The multipliers then follow in closed form,
\begin{equation}
    \lambda_{\mathbf{q},n}
    = \frac{1}{2N}\left[\frac{1}{S(\mathbf{q},\omega_n)}
    -\frac{1}{S_0(\mathbf{q},\omega_n)}\right],
\label{eq:lambda}
\end{equation}
determined entirely by the deviation of the measured correlation function from its non-interacting reference. In the End Matter, we discuss their physical interpretation as an effective interaction. The relative entropy then becomes
\begin{equation}
    H_{\text{rel}} \approx \frac{1}{2}\sum_{\mathbf{q}\neq0,n}\left[\ln\left(\frac{S(\mathbf{q},\omega_n)}{S_0(\mathbf{q},\omega_n)}\right) -\frac{S(\mathbf{q},\omega_n)}{S_0(\mathbf{q},\omega_n)} +1\right].
\label{eq:H_rel_main}
\end{equation}
\Cref{eq:H_rel_main} has the form of \cref{eq:classic_entropy_estimation}, with the ratio $S(\mathbf q)/S_0(\mathbf q)$ replaced, mode by mode, by $S(\mathbf q,\omega_n)/S_0(\mathbf q,\omega_n)$. \Cref{eq:H_rel_main} requires only these two correlation functions, without the interaction potential, and applies unchanged to bosons and fermions. As in the classical case, the Gaussian approximation does not retain the variational property of the exact maximum-entropy solution, so this closed-form result is not a proven bound.

The classical result~\cite{ariel2020inferring} is recovered as a limiting case of \cref{eq:H_rel_main}. In the classical limit the paths shrink to points, and their remaining variation along imaginary time is that of free particles. The terms with $n\neq0$ then vanish, since $S(\mathbf q,\omega_n)/S_0(\mathbf q,\omega_n)\to1$, while the $n=0$ term takes its classical value, $S(\mathbf q,0)/S_0(\mathbf q,0)\to S(\mathbf q)$ (Supplemental Material, \cref{SI:high_frequency}). The sum therefore reduces to \cref{eq:classic_entropy_estimation}.

\textit{Excess entropy and free energy.}---Quantum mechanically, the excess entropy $H_{\text{ex}}\equiv H-H_0$ differs from the negative relative entropy, but we show that the difference is fixed by the same two correlation functions (End Matter, \cref{sec:legendre}). For a non-interacting reference, the two differ by the interaction-induced change in kinetic energy, $H_{\text{ex}} = H_{\text{rel}} + \beta\big(E_{\text{kin}}-E_{\text{kin}}^{0}\big)$, where $E_{\text{kin}}$ and $E_{\text{kin}}^{0}$ are the kinetic energies of the interacting state and of the non-interacting reference, which are equal in the classical limit. We define an effective excess free energy function $F_{\text{ex}}$ and show that $E_{\text{kin}}-E_{\text{kin}}^{0}=-m(\partial F_{\text{ex}}/\partial m)_\lambda$, where $m$ is the particle mass. This relation identifies $F_{\text{ex}}$ as the excess free energy of the inferred state, similar to path-integral calculations of isotope effects~\cite{ceriotti2013efficient,marsalek2014efficient,perez2011path,vanivcek2005quantum}. Its closed form is minus the logarithmic term of \cref{eq:H_rel_main},
\begin{equation}
    \beta F_{\text{ex}}\approx \frac{1}{2}\sum_{\mathbf{q}\neq0,n}
    \ln\frac{S_{0}(\mathbf{q},\omega_{n})}{S(\mathbf{q},\omega_{n})} ,
\label{eq:Fex}
\end{equation}
and its Legendre transform at fixed multipliers gives the excess entropy,
\begin{equation}
    H_{\text{ex}}\approx \frac{1}{2}\sum_{\mathbf{q}\neq0,n}
    \left[\ln\frac{S}{S_{0}}
    -\beta\,\frac{S-S_{0}}{S_{0}^{2}}\,
    \frac{\partial S_{0}}{\partial\beta}\right] ,
\label{eq:Hex_explicit}
\end{equation}
where $S$ and $S_0$ stand for $S(\mathbf q,\omega_n)$ and $S_0(\mathbf q,\omega_n)$, and $\partial S_0/\partial\beta$ is taken at fixed $\mathbf{q}$ and $n$.

The correlation free energy, $\Delta F_{\text{corr}}$, excludes the first-order correction of $F_{\text{ex}}$, evaluated in the non-interacting reference. It is minus the relative entropy with its arguments reversed, $\Delta F_{\text{corr}}=-\beta^{-1}D(\hat P_0\|\hat P)$, a general relation for two thermal states that, as we show in \cref{SI:reversed} of the Supplemental Material, here evaluates to
\begin{equation}
    \beta\Delta F_{\text{corr}} \approx \frac{1}{2}\sum_{\mathbf q\neq0,n}
    \left[\ln\frac{S_0}{S}-\frac{S_0}{S}+1\right] .
\label{eq:Fcorr}
\end{equation}
When a system can be described with the random-phase approximation (RPA)~\cite{gell1957correlation,fetter2012quantum,altland2010condensed}, \cref{eq:Fcorr} reduces to the correlation free energy of the RPA. However, \cref{eq:Fex,eq:Hex_explicit,eq:Fcorr} do not require the interaction potential or an integration over the coupling constant. %

\begin{figure}
  \centering
  \includegraphics{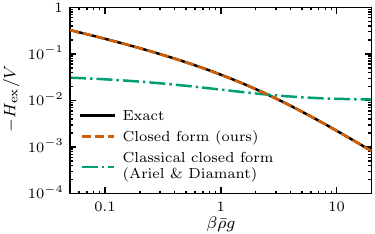}
  \caption{%
    Minus the excess entropy per volume of the Bogoliubov gas, in units of
    $(2m\bar\rho g/\hbar^2)^{3/2}$, versus $\beta\bar\rho g$, with $\bar\rho=N/V$ the mean density and $g$ the coupling constant. Solid line: exact. Dashed line: closed form,
    \cref{eq:Hex_explicit}. Dash-dotted line: \cref{eq:classic_entropy_estimation}
    evaluated with the equal-time ratio $S(q)/S_0(q)$.%
  }
  \label{fig:bose}
\end{figure}

\textit{Static structure factor as input.}---When $S(\mathbf q,\omega_n)$ is not available, it is possible to reconstruct it from $S(\mathbf q)$ and $S_0(\mathbf q,\omega_n)$ if the interaction is instantaneous. The exact multipliers then do not depend on $n$, $\lambda_{\mathbf q,n}=\lambda_{\mathbf q}$ (Supplemental Material, \cref{SI:n_independent}). Then $\lambda_{\mathbf q}$ in \cref{eq:lambda} is fixed by requiring $\sum_nS(\mathbf q,\omega_n)=\beta S(\mathbf q)$, i.e.,
\begin{equation}
    S(\mathbf q)=\frac{1}{\beta}\sum_n\frac{S_0(\mathbf q,\omega_n)}{1+2N\lambda_{\mathbf q}S_0(\mathbf q,\omega_n)} .
\label{eq:S_reconstructed}
\end{equation}
\Cref{eq:S_reconstructed} has a single solution $\lambda_{\mathbf q}$ for which every term of the sum is positive. \Cref{eq:H_rel_main,eq:Fex,eq:Hex_explicit,eq:Fcorr} can then be evaluated with the reconstructed $S(\mathbf q,\omega_n)$ from \cref{eq:lambda}. This route still needs $S_0(\mathbf q,\omega_n)$, which supplies the dependence of $S(\mathbf q,\omega_n)$ on $\omega_n$.

\textit{Bogoliubov gas.}---For a weakly interacting Bose gas within the Bogoliubov approximation, our framework reproduces the entropy exactly, while the classical relation evaluated with equal-time correlations does not (\cref{fig:bose}). In this approximation, the quasiparticles, like the particles of the ideal reference, are independent harmonic oscillators, so the density fluctuations are Gaussian. The three- and higher-point correlations neglected in the closed forms then vanish, so an exact agreement is expected. \Cref{eq:lambda} returns the coupling constant of the microscopic Hamiltonian, and \cref{eq:Fex,eq:Hex_explicit} give the free energy and entropy of the quasiparticles from the correlation functions alone (Supplemental Material, \cref{sec:harmonic}). 

\begin{figure*}
\includegraphics[width=\linewidth]{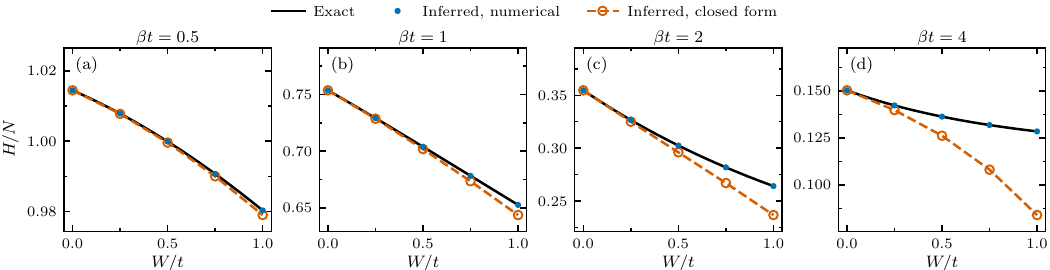}
\caption{Benchmark for spinless fermions on a ring, \cref{eq:tW_model}, at
$L=12$, $N=6$. Entropy per particle $H/N$ against interaction
strength $W/t$ at (a) $\beta t=0.5$, (b) $\beta t=1$, (c) $\beta t=2$, and
(d) $\beta t=4$, effectively zero temperature for this system; the vertical scales differ between panels. The solid line is
exact diagonalization. Dots are numerical inference without the Gaussian
approximation, which reproduces the exact entropy at
every point. Circles are the closed-form
estimate, \cref{eq:H_rel_main}, converted to $H$ with the exact kinetic energy. Different ring sizes are in the Supplemental Material (\cref{fig:SI_ring_size}).}
\label{fig:benchmark}
\end{figure*}

\textit{Spinless fermions on a ring.}---We now test the closed-form relation on a small interacting system, whose density modes are far from Gaussian. We consider $N$ spinless fermions on a ring of $L$ sites with unit lattice spacing and periodic boundary conditions,
\begin{equation}
    \hat{\mathcal{H}}
    = -t\sum_{j=1}^{L}\left(\hat{c}_j^{\dagger}\hat{c}_{j+1}+\mathrm{h.c.}\right)
    + W\sum_{j=1}^{L}\hat{n}_j\hat{n}_{j+1},
\label{eq:tW_model}
\end{equation}
where $\hat{c}_j$ annihilates a fermion on site $j$ and $\hat{n}_j=\hat{c}_j^{\dagger}\hat{c}_j$. Exact diagonalization provides the exact entropy, the kinetic energy, and the full density-density correlation function $S(q,\omega_n)$. We convert \cref{eq:H_rel_main} to the total entropy with the exact kinetic energy (End Matter, \cref{sec:lattice}). Because all inputs are exact, deviations from the exact entropy represent the error of the closed form (\cref{fig:benchmark}).

A numerical inference that uses our effective partition function without the Gaussian approximation reproduces the exact entropy here (dots in \cref{fig:benchmark}). The bound of the exact inference is thus very tight in this case (Supplemental Material, \cref{SI:control}). %
The closed-form entropy is accurate at weak coupling and high temperature, and deviates increasingly as the coupling grows or the temperature drops (\cref{fig:benchmark}). At $L=12$, $N=6$, and $W=t$, the relative error in $H$ grows from $0.13\%$ at $\beta t=0.5$ to $35\%$ at $\beta t=4$, which is effectively zero temperature for this system, because its lowest excitation energy is at least four times the thermal energy. This deviation on cooling comes from the higher-order density cumulants neglected by the Gaussian approximation. The approximation also does not retain the bound of the exact inference, so the closed-form entropy can lie below the exact entropy, as it does at every temperature and coupling examined. In the Supplemental Material, we show results for different ring sizes (\cref{fig:SI_ring_size}).

\begin{figure}
  \centering
  \includegraphics{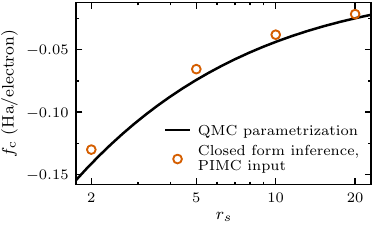}
  \caption{%
    Correlation free energy of the spin-unpolarized uniform
    electron gas at $\Theta = 1$, in units of Hartree (Ha) per electron. Circles: \cref{eq:Fcorr} evaluated on
    the path-integral
    Monte Carlo (PIMC) imaginary-time correlation functions of Refs.~\cite{chuna2025estimates,chuna_thomas_michael_2025_3923} for $N=34$
    electrons, corrected to the thermodynamic limit within the RPA; statistical errors are smaller than the symbols. Solid line: $f_{\rm c} = f_{\rm xc} - f_{\rm x}\equiv\Delta F_{\text{corr}}/N$, with
    $f_{\rm xc}\equiv F_{\text{ex}}/N$ from the corrected quantum Monte Carlo (QMC) parametrization of Refs.~\cite{karasiev2014accurate,karasiev2018nonempirical} and $f_{\rm x}$
    the exact finite-temperature exchange free energy.%
  }
  \label{fig:fc}
\end{figure}

\textit{Warm dense electron gas.}---We now apply \cref{eq:Fcorr} to a continuum system for which the answer is known independently: the spin-unpolarized uniform electron gas~\cite{giuliani2005quantum,loos2016uniform,dornheim2018uniform}. Its state is fixed by the density parameter $r_s = \bar{a}/a_{\rm B}$, with $\bar{a}$ the Wigner--Seitz radius and $a_{\rm B}$ the Bohr radius, and by the reduced temperature $\Theta = k_{\rm B}T/E_{\rm F}$, with $E_{\rm F}=\hbar^2k_{\rm F}^2/2m$ the Fermi energy and $k_{\rm F}$ the Fermi wave number. We work at $\Theta=1$ and $2 \le r_s \le 20$, which spans moderate to strong coupling. The neutralizing background cancels the Hartree term, so the free-energy difference $F_{\text{ex}}$ per particle is the exchange-correlation free energy $f_{\rm xc}$. Its first-order part is the exchange free energy $f_{\rm x}$, known exactly at finite temperature~\cite{horovitz1974exchange}, and the remainder is the correlation free energy per particle $f_{\rm c} = f_{\rm xc} - f_{\rm x}$, which we compare with the closed form in \cref{eq:Fcorr}, taking $f_{\rm xc}$ from the corrected quantum Monte Carlo-based parametrization of Karasiev \emph{et al.}~\cite{karasiev2014accurate,karasiev2018nonempirical} (solid line in \cref{fig:fc}; Supplemental Material, \cref{SI:ueg_reference}). 

Our input is the imaginary-time density correlation function $\langle\rho_{\mathbf q}^{(\tau)}\rho_{-\mathbf q}^{(0)}\rangle/N$, computed by \citet{chuna2025estimates} with path-integral Monte Carlo for $N=34$ electrons~\cite{chuna_thomas_michael_2025_3923}. We transform it to $S(\mathbf q,\omega_n)$ with \cref{eq:Matsubara_density-density}, take $S_0$ as that of the ideal Fermi gas, and obtain the statistical errors from the $1000$ independent simulations. The sum over $\mathbf q$ in \cref{eq:Fcorr} runs up to the largest simulated wave number, $6.8k_{\rm F}$, where it still grows by at most $0.35\%$ of $|f_{\rm c}|$ per $k_{\rm F}$ (Supplemental Material, \cref{sec:ueg}).

The reference describes the thermodynamic limit, so we correct our result for the finite size of the simulation cell. The cell has no density modes below $0.63k_{\rm F}$, which carry most of $f_{\rm c}$ in the thermodynamic limit, because the Coulomb interaction is strongest at small $q$. As in finite-size corrections of quantum Monte Carlo data for the electron gas~\cite{chiesa2006finite,dornheim2016ab}, we treat these modes in the RPA, which becomes exact at long wavelengths. This correction is between $49\%$ and $68\%$ of each plotted value, but it is insensitive to the choice of model. Replacing the RPA by a simple model, a static local field correction as in dielectric theories~\cite{singwi1968electron,tanaka1986thermodynamics,dornheim2019static}, changes $f_{\rm c}$ by about $1\%$ (Supplemental Material, \cref{SI:ueg_finite_size}).

The closed form follows the reference over the whole range of $r_s$ (\cref{fig:fc}). It underestimates the magnitude of $f_{\rm c}$ by $8.0\%$ at $r_s = 2$ and by $13.3\%$ at $r_s = 20$, with statistical errors of $1.2\%$ at $r_s=2$ and below $0.1\%$ at larger $r_s$. Since $f_{\rm x}$ is exact, this corresponds to an error of $4.9\%$ in $f_{\rm xc}$ at $r_s = 2$ and $9.8\%$ at $r_s = 20$. To check whether this deviation comes from the input, we reconstruct $S(\mathbf q,\omega_n)$ from the simulated $S(\mathbf q)$ with \cref{eq:S_reconstructed}, and $f_{\rm c}$ moves by no more than about $1\%$ (Supplemental Material, \cref{SI:ueg_reconstruction}). We therefore attribute the deviation mainly to the Gaussian approximation.

\textit{Conclusions.}---We have shown that imaginary-time density correlations contain the thermodynamic information that equal-time structure misses, and that with them the structure--entropy relation of \citet{ariel2020inferring} carries over to quantum systems. The closed forms recover the classical relation and the RPA as special cases and reproduce the Bogoliubov entropy exactly. On the ring, the closed form is accurate at high temperature and deviates on cooling; its errors are one-sided in every case examined, although no bound is proven. For the warm dense electron gas, with simulated imaginary-time correlation functions as input, it underestimates the magnitude of the correlation free energy by up to $13\%$. This deviation grows with coupling, as does the error of the Gaussian approximation on the ring (Supplemental Material, \cref{SI:cancellation}). Unlike its classical counterpart, the construction is derived for thermal equilibrium. Out of equilibrium it could be evaluated at an effective temperature, but this has not been tested.

Let us summarize the advantages of \cref{eq:Fcorr}. It requires a single simulation or experiment at the state point, whereas thermodynamic integration requires the interaction energy at every coupling between zero and the physical value. For the electron gas, this integration is costly, because the sign problem of path-integral Monte Carlo worsens as the coupling weakens~\cite{chuna2025second}. An extended-ensemble method is exact at a single state point but requires dedicated simulations with scaled interactions~\cite{dornheim2025direct}; \cref{eq:Fcorr} only post-processes the density correlations of standard simulations or experiments, without the interaction potential. Accurate free energies already exist for the electron gas, but few for warm dense hydrogen~\cite{dornheim2025direct}, whose imaginary-time correlation functions have been computed~\cite{dornheim2024abinitio}.

Another advantage of our framework is that, for an instantaneous interaction, $S(\mathbf q,\omega_n)$ can be obtained from $S(\mathbf q)$ and $S_0(\mathbf q,\omega_n)$ via \cref{eq:lambda,eq:S_reconstructed}. In that case the information that $S(\mathbf q)$ lacks is reconstructed from $S_0(\mathbf q,\omega_n)$. Indeed, in all three benchmarks, where the interaction is instantaneous, $S(\mathbf q)$ gives nearly the same results as $S(\mathbf q,\omega_n)$.

Our approach of imaginary-time paths can also be used to adapt entropy estimators based on compression~\cite{martiniani2019quantifying,avinery2019universal} or mutual information~\cite{nir2020machine,benshimon2026machine} to quantum systems. Simulations at additional particle numbers would replace the finite-size correction with an extrapolation. Simulated~\cite{dornheim2023physical,tolias2024fourier} or measured~\cite{dornheim2022accurate,dornheim2024xray} imaginary-time density correlations thus offer a route to quantum entropies and free energies where coupling paths are costly or, as in experiments, unavailable.

\begin{acknowledgments}
\textit{Acknowledgments.}---S.M. acknowledges support from the Tel Aviv University Center for Artificial Intelligence and Data Science (TAD) and from the Israeli Planning and Budgeting Committee (VATAT). S.M. would like to thank Haim Diamant, Barak Hirshberg, Michael Galperin, and Gil Ariel for fruitful discussions and suggestions. S.M. used Claude Opus 4.8 and 5.5, Claude Fable 5.2, and GPT-5.6 to assist with coding and writing, and reviewed all of their output.
\end{acknowledgments}

\textit{Data availability.}---The code that produces the results and figures of this Letter is available at \url{https://github.com/SagiMeir/imaginary-time-entropy}; the path-integral Monte Carlo input data are available from Ref.~\cite{chuna_thomas_michael_2025_3923}.

\bibliography{refs}

\appendix
\section*{End Matter}

\section{Derivation of the closed form}
\label{sec:variational}
\label{sec:gaussian}

We derive \cref{eq:lambda,eq:H_rel_main} and show that they need a single approximation, the Gaussian approximation of the classical derivation~\cite{ariel2020inferring}. We start by defining the measure
\begin{equation}
    \oint\mathcal{D}\mathbf{R} \equiv \frac{1}{N!}\sum_{\sigma}\eta^{P_\sigma}\oint_{\mathbf r_j^{(\beta\hbar)}=\mathbf r_{\sigma(j)}^{(0)}}\prod_{j=1}^N\mathcal{D}\mathbf{r}_j^{(\tau)} ,
\label{eq:path_measure}
\end{equation}
which sums over particle permutations $\sigma$~\cite{ceperley1995path,hirshberg2019path}, where $P_\sigma$ is the number of transpositions in $\sigma$ and $\eta=+1$ for bosons and $-1$ for fermions; for Boltzmann statistics it reduces to the product of independent closed paths. Quantum statistics enter only through this measure, so the derivation holds for bosons and fermions alike. 
We write $S(\mathbf q,\omega_n)$ symmetrically in terms of the Fourier--Matsubara density modes of a set of paths $R=\{\mathbf r_j^{(\tau)}\}$,
\begin{equation}
    \rho_{\mathbf{q},n}[R] = \frac{1}{\sqrt{\beta}\hbar}\int_0^{\beta\hbar}\mathrm{d}\tau\, e^{i\omega_n\tau}\sum_{j} e^{-i\mathbf{q}\cdot\mathbf{r}_j^{(\tau)}} ,
\label{eq:path_modes}
\end{equation}
where $NS(\mathbf q,\omega_n)=\langle|\rho_{\mathbf{q},n}[R]|^2\rangle$ (\cref{SI:density_correlations}).

We now express the relative entropy, \cref{eq:H_rel_def}, in the position representation. The reference state is $\hat{P}_0 = e^{-\beta\hat{\mathcal{H}}_0}/\mathcal{I}[0]$, where $\hat{\mathcal{H}}_0$ is the Hamiltonian of the non-interacting gas and $\mathcal{I}[0]=\text{tr}\,e^{-\beta\hat{\mathcal{H}}_0}$ is the ideal quantum-gas partition function. In this representation, $\mathcal{I}[0]=\oint\mathcal{D}\mathbf{R}\,e^{-\mathcal{A}_0[R]}$, with $\mathcal{A}_0[R]\equiv\hbar^{-1}\int_0^{\beta\hbar}\mathrm{d}\tau\sum_{j} \tfrac{1}{2}m|\dot{\mathbf{r}}_j^{(\tau)}|^2$ the free-particle Euclidean action, so each set of paths has the reference weight $P_0[R]=e^{-\mathcal{A}_0[R]}/\mathcal{I}[0]$. We then extremize the relative entropy of the path weights $P[R]$ with the Lagrangian (\cref{SI:maximization})
\begin{multline}
    \mathcal{L}=-\oint\mathcal{D}\mathbf{R}\,P\ln\frac{P}{P_0}
    -\mu\Big(\oint\mathcal{D}\mathbf{R}\,P-1\Big)\\
    -\sum_{\mathbf{q},n}\lambda_{\mathbf{q},n}\Big(\oint\mathcal{D}\mathbf{R}\,P\,|\rho_{\mathbf{q},n}|^2-NS(\mathbf{q},\omega_n)\Big),
\label{eq:Lagrangian}
\end{multline}
where $\lambda_{\mathbf{q},n}$ enforces the given $S(\mathbf q,\omega_n)$ at each $\mathbf{q}\neq0$ (all sums over $\mathbf q$ exclude $\mathbf q=0$) and $\mu$ the normalization. Setting $\delta\mathcal{L}/\delta P[R]=0$ reweights the reference weight $P_0[R]$ of each path by $\exp(-\sum_{\mathbf{q},n}\lambda_{\mathbf{q},n}|\rho_{\mathbf{q},n}[R]|^2)$, with normalization
\begin{multline}
    \mathcal{Z}=e^{N\sum_{\mathbf{q},n}\lambda_{\mathbf{q},n}S(\mathbf{q},\omega_n)}\\
    \times\oint\mathcal{D}\mathbf{R}\,
    \exp\Big(-\mathcal{A}_0[R]-\sum_{\mathbf{q},n}\lambda_{\mathbf{q},n}
    \big|\rho_{\mathbf{q},n}[R]\big|^2\Big),
\label{eq:Zpath}
\end{multline}
and since the constraints hold, the relative entropy at the solution is $H_{\text{rel}} = \ln\mathcal{Z} - \ln\mathcal{I}[0]$.

As in the classical derivation~\cite[Appendix~A]{ariel2020inferring}, collective variables $\tilde{\rho}_{\mathbf{q},n}$, introduced through a delta function with a conjugate field $\tilde{\psi}_{\mathbf{q},n}$, carry the quadratic term; integrating them out leaves the paths coupled to $\tilde{\psi}$ only linearly (\cref{SI:field}):
\begin{multline}
    \mathcal{Z}\propto
    e^{N\sum_{\mathbf{q},n}\lambda_{\mathbf{q},n}S(\mathbf{q},\omega_n)}
    \Big(\prod_{\mathbf{q},n}\lambda_{\mathbf{q},n}^{-1/2}\Big)\\
    \times\int\mathcal{D}\tilde{\psi}\;
    e^{-\mathcal{H}_\psi[\tilde{\psi}]},
\label{eq:Zaux}
\end{multline}
with $\mathcal{H}_\psi=\sum_{\mathbf{q},n}|\tilde{\psi}_{\mathbf{q},n}|^2/4\lambda_{\mathbf{q},n}-\ln\mathcal{I}[\tilde{\psi}]$ and
\begin{equation}
    \mathcal{I}[\tilde{\psi}]\equiv
    \oint\mathcal{D}\mathbf{R}\,
    \exp\Big(-\mathcal{A}_0[R]-i\sum_{\mathbf{q},n}\tilde{\psi}_{\mathbf{q},n}\rho_{\mathbf{q},n}[R]\Big).
\label{eq:Hpsi}
\end{equation}
For $\lambda_{\mathbf{q},n}<0$, \cref{eq:Zaux} is defined by analytic continuation from $\lambda_{\mathbf{q},n}>0$ (\cref{SI:field}). Everything is exact up to this step. 

The Gaussian approximation is to expand $\ln\mathcal{I}[\tilde{\psi}]$ to quadratic order in the conjugate field; the neglected terms are of higher order in $\tilde{\psi}$, set by the three- and higher-point density correlations of the reference. The first-order term, $-i\langle\rho_{\mathbf{q},n}\rangle_{\mathcal{A}_0}$, vanishes for $\mathbf{q}\neq 0$ by translational invariance of the free gas, and translational and imaginary-time-translational invariance make the second-order term diagonal, with coefficient $-N\,S_0(\mathbf{q},\omega_n)$ (\cref{SI:Approximation}):
\begin{equation}
    \ln\mathcal{I}[\tilde{\psi}]\approx\ln\mathcal{I}[0]
    -\frac{N}{2}\sum_{\mathbf{q},n}S_0(\mathbf{q},\omega_n)\,\big|\tilde{\psi}_{\mathbf{q},n}\big|^2 .
\label{eq:secondorder}
\end{equation}
The field integral in \cref{eq:Zaux} is then elementary; carrying it out cancels the field-space constants and gives
\begin{equation}
    H_{\text{rel}}(\lambda)
    \approx \sum_{\mathbf{q},n}\Big[N\lambda_{\mathbf{q},n}S
    -\frac{1}{2}\ln\big(1+2N\lambda_{\mathbf{q},n}S_0\big)\Big],
\label{eq:Hrel_lambda}
\end{equation}
with $S$ and $S_0$ short for $S(\mathbf{q},\omega_n)$ and $S_0(\mathbf{q},\omega_n)$. Setting $\partial H_{\text{rel}}/\partial\lambda_{\mathbf{q},n}=0$ gives \cref{eq:lambda}, and substituting it gives \cref{eq:H_rel_main}. %

\section{Excess entropy and the kinetic shift}
\label{sec:legendre}

Using \cref{eq:H_rel_def} with the thermal reference, the relative entropy differs from the excess entropy by the change in kinetic energy, $H_{\text{rel}} = H_{\text{ex}} + \beta(E_{\text{kin}}^{0}-E_{\text{kin}})$, where $E_{\text{kin}}=\text{tr}(\hat{P}\hat{\mathcal{H}}_0)$ and
$E_{\text{kin}}^{0}=\text{tr}(\hat{P}_0\hat{\mathcal{H}}_0)$ are the kinetic energies of the interacting and reference states, which differ quantum mechanically because $\hat{\mathcal{H}}_0$ does not commute with the potential energy. However, we can obtain them from the correlation function. Note that the mass enters only through the free-particle weight, $e^{-\mathcal{A}_0}$, and the normalization of the measure. The density mode $\rho_{\mathbf{q},n}[R]$ is a functional of the paths alone: two particles of different mass tracing the same path return the same value. Therefore, the kinetic energies are mass derivatives at fixed multipliers, $E_{\text{kin}}^{0}=\beta^{-1}m\partial_m \ln\mathcal{I}[0]$ and $E_{\text{kin}}=\beta^{-1}m\left(\partial_m\ln\mathcal{Z}_{\lambda}\right)_\lambda$, where $\mathcal{Z}_{\lambda} \equiv \mathcal{Z}\, e^{-N\sum_{\mathbf{q},n}\lambda_{\mathbf{q},n}S(\mathbf{q},\omega_{n})}$ is the path integral of \cref{eq:Zpath}. 

By defining
\begin{equation}
    F_{\text{ex}}\equiv -\beta^{-1}\ln\left(\mathcal{Z}_{\lambda}/\mathcal{I}[0]\right) ,
\label{eq:Fex_def}
\end{equation}
we obtain the following relation,
\begin{equation}
    E_{\text{kin}}-E_{\text{kin}}^{0}
    = -m\left(\frac{\partial F_{\text{ex}}}{\partial m}\right)_\lambda ,
\label{eq:kinetic_shift_mass}
\end{equation}
which identifies $F_{\text{ex}}$ as the excess free energy of the inferred state~\cite{ceriotti2013efficient,marsalek2014efficient,perez2011path,vanivcek2005quantum}. Comparing the definition of $\mathcal{Z}_{\lambda}$ with \cref{eq:Hrel_lambda,eq:lambda}, using $1+2N\lambda_{\mathbf{q},n}S_{0}=S_{0}/S$, gives \cref{eq:Fex}. Differentiating \cref{eq:Fex} at fixed $\lambda_{\mathbf{q},n}$, where $S$ varies with the mass through $S_{0}$ according to \cref{eq:lambda}, gives the kinetic shift
\begin{equation}
    \beta\left(E_{\text{kin}}-E_{\text{kin}}^{0}\right)
    \approx \frac{1}{2}\sum_{\mathbf{q},n}
      \frac{S-S_{0}}{S_{0}^{2}}\;
      m\frac{\partial S_{0}}{\partial m} ,
\label{eq:kinetic_shift_explicit}
\end{equation}
so the shift is set by the same two correlation functions that determine
$H_{\text{rel}}$, together with the mass dependence of the reference, which can be traded for a temperature dependence (\cref{SI:legendre}). The Legendre transform $H_{\text{ex}} = -\beta F_{\text{ex}} + \beta\,(\partial (\beta F_{\text{ex}})/\partial \beta)_\lambda$ then gives \cref{eq:Hex_explicit}. The derivative $\partial S_0/\partial\beta$ at fixed $\mathbf{q}$ and $n$ includes the dependence of $\omega_n=2\pi n/\beta\hbar$ on $\beta$. In the classical limit only the $n=0$ mode contributes, \cref{eq:kinetic_shift_explicit} gives $E_{\text{kin}}=E_{\text{kin}}^{0}$, and \cref{eq:Hex_explicit} reduces to \cref{eq:classic_entropy_estimation}.

\section{Effective interaction and the random-phase approximation}
\label{sec:rpa}

The multipliers give the effective interaction that reproduces the given correlations in RPA form. Writing the response functions as $\chi=-\bar\rho\,S$ and $\chi_0=-\bar\rho\,S_0$ with $\bar\rho=N/V$, \cref{eq:lambda} becomes
\begin{equation}
\begin{split}
    2V\lambda_{\mathbf q,n}
    &= \frac{1}{\chi_0(\mathbf q,i\omega_n)}-\frac{1}{\chi(\mathbf q,i\omega_n)}\\
    &= U_{\mathbf q}\left[1-G(\mathbf q,i\omega_n)\right] ,
\end{split}
\label{eq:U_lambda_lfc}
\end{equation}
where the second equality defines the local field correction $G$ for a pair interaction $U_{\mathbf q}$~\cite{ichimaru1982strongly,giuliani2005quantum}. Classically, at $n=0$, the first equality reduces to the multiplier of \citet{ariel2020inferring}, which is proportional to minus the direct correlation function of the Ornstein--Zernike relation~\cite{hansen2006theory} and approaches the pair potential at weak coupling.

The RPA sets $G=0$, so $2V\lambda_{\mathbf q,n}=U_{\mathbf q}$ in \cref{eq:U_lambda_lfc} and $S_0/S=1-U_{\mathbf q}\chi_0$; \cref{eq:Fcorr} then becomes the RPA correlation free energy~\cite{gell1957correlation,fetter2012quantum,altland2010condensed},
\begin{equation} %
    \beta\Delta F_{\text{corr}}^{\text{RPA}} = \frac{1}{2}\sum_{\mathbf q\neq0,n}
    \left[\ln\big(1-U_{\mathbf q}\chi_0\big) + U_{\mathbf q}\chi_0\right] .
\label{eq:F_corr_rpa}
\end{equation}

\section{Spinless-fermion ring: interaction and rescaled reference}
\label{sec:lattice}

On the ring, $\hat{\rho}_q=\sum_j e^{-iqj}\hat{n}_j$ with $q=2\pi k/L$, $k=1,\dots,L-1$, and the interaction is a pair-density coupling. Using $\hat{\rho}_0=N$,
\begin{equation}
    \hat{U} = W\sum_j\hat{n}_j\hat{n}_{j+1}
    = \frac{1}{2L}\sum_{q\neq0}U_q\,\hat{\rho}_q\hat{\rho}_{-q}
    + \frac{WN^2}{L},
\label{eq:U_lattice}
\end{equation}
with $U_q = 2W\cos q$ and the volume of \cref{eq:U_lambda_lfc} replaced by the number of sites $L$. The
constant $WN^2/L$ shifts only absolute free energies. With $2L\lambda_{q,n}=U_q$, the weight in \cref{eq:Zpath} is that of the exact $\hat{P}$, so the numerical inference recovers $\hat{P}$ (dots in \cref{fig:benchmark}).

To convert \cref{eq:H_rel_main} to the total entropy, we use a rescaled reference. The leading high-frequency coefficient of the density-density correlation function is fixed by the kinetic energy (\cref{SI:sumrule}),
\begin{equation}
    \omega_n^2\,S(q,\omega_n)\;\xrightarrow{|n|\to\infty}\;
    \frac{2(1-\cos q)}{N\hbar^2}\,\big|E_{\mathrm{kin}}\big|.
\label{eq:tail_lattice}
\end{equation}
The interacting system and the
physical free reference have different kinetic energies, so the ratio $S/S_0$ does not approach unity at large $|n|$,
and the Matsubara sum in \cref{eq:H_rel_main} does not converge directly
against the physical reference. We resolve this by evaluating the closed
form against a free reference of rescaled kinetic scale,
\begin{equation}
    \hat{P}_0^{\gamma} = e^{-\beta\gamma\hat{\mathcal{H}}_0}/\mathcal{I}[0;\gamma],
\label{eq:P0gamma}
\end{equation}
with the scale $\gamma^{\ast}$ fixed by matching the kinetic energy of the
scaled reference, $E_{\mathrm{kin}}^{0}(\gamma)\equiv\operatorname{tr}(\hat{P}_0^{\gamma}\,\gamma\hat{\mathcal{H}}_0)$, to that of the interacting state,
\begin{equation}
    E_{\mathrm{kin}}^{0}(\gamma^{\ast}) = E_{\mathrm{kin}},
\label{eq:gammastar}
\end{equation}
which ensures that $S/S_0^{\gamma}\to1$ at large $|n|$. The result is transformed exactly back to the physical reference at $\gamma=1$ (\cref{SI:bridge}), giving the estimator $H_{\text{est}}=H_{\text{rel}}^{\text{Gauss}}(\gamma^{\ast})+\beta\gamma^{\ast}E_{\text{kin}}+\ln\mathcal{I}[0;\gamma^{\ast}]$, with \cref{eq:H_rel_main} against $S_0^{\gamma^{\ast}}$ and the exact $E_{\text{kin}}$. Repeated against this reference, the inference shown by the dots in \cref{fig:benchmark} still reproduces the exact entropy to within $0.07\%$ (\cref{SI:bridge}), so the deviation on cooling comes mainly from the Gaussian approximation.

\clearpage
\onecolumngrid
\addtocontents{toc}{\protect\def\protect\showappendix{1}} %

\renewcommand{\thefigure}{S\arabic{figure}}
\renewcommand{\thetable}{S\arabic{table}}
\renewcommand{\theequation}{S\arabic{equation}}
\setcounter{figure}{0} 
\setcounter{table}{0}
\setcounter{equation}{0}
\counterwithout*{equation}{section} %
\renewcommand{\thesubsection}{\arabic{subsection}}
\setcounter{subsection}{0}
\setcounter{section}{0} %
\renewcommand{\thesection}{S\arabic{section}}
\makeatletter
\def\p@subsection{\thesection.} %
\def\theHsection{S\arabic{section}} %
\def\theHsubsection{\theHsection.\arabic{subsection}}
\def\theHfigure{S\arabic{figure}}\def\theHtable{S\arabic{table}} %
\def\@sectioncntformat#1{\csname the#1\endcsname} %
\makeatother
\crefalias{appendix}{smsection}\crefalias{subappendix}{smsection} %
\newcommand{\smfootnote}[1]{\footnote{\makebox[0pt][l]{\parbox[t]{\dimexpr\textwidth-1.5em\relax}{#1}}}} %

\begin{center}
    \begin{large}
        \textbf{Supplemental Material}
    \end{large}
\end{center}

\makeatletter

\def\showappendix{0} %
\let\oldcontentsline\contentsline

\renewcommand{\contentsline}[4]{%
  \ifnum\showappendix=1
    \oldcontentsline{#1}{#2}{#3}{#4}%
  \fi
}

\begingroup
\let\appendix\relax 
\@starttoc{toc} %
\endgroup

\let\contentsline\oldcontentsline %
\vspace{1cm} %
\makeatother

\textit{Conventions.}---The following conventions hold throughout the Supplemental Material: $N$ particles of mass $m$ occupy a volume $V$ in $d$ dimensions with periodic boundary conditions, $\bar\rho=N/V$, $\beta=1/k_{\mathrm B}T$, and $\hbar$ is kept explicit. Wave vectors satisfy $\mathbf q\neq0$ unless stated otherwise, $\epsilon_q=\hbar^2q^2/2m$, and $\sum_{\mathbf q,n}$ runs over $\mathbf q\neq0$ and all integers $n$. The correlation functions $S(\mathbf q,\omega_n)$ have units of inverse energy, and the multipliers $\lambda_{\mathbf q,n}$ units of energy. In spectral sums $|m\rangle$ and $|l\rangle$ are eigenstates of the Hamiltonian $\hat{\mathcal H}$; this $m$ is a summation index, unrelated to the particle mass.

\section{Density correlations in imaginary time} 
\label{SI:density_correlations} 
We collect here the definitions of the density correlation functions used throughout the paper and clarify the relation between their equal-time, imaginary-time, and Matsubara representations.

All the correlation functions below are built from the Fourier components of the particle density. The microscopic density operator is 
\begin{equation} 
\hat{\rho}(\mathbf r) = \sum_{j=1}^{N} \delta\left(\mathbf r-\hat{\mathbf r}_j\right), \end{equation} 
with Fourier components 
\begin{equation} 
\hat{\rho}_{\mathbf q} = \int \mathrm d\mathbf r\, e^{-i\mathbf q\cdot\mathbf r}\hat{\rho}(\mathbf r) = \sum_{j=1}^{N} e^{-i\mathbf q\cdot\hat{\mathbf r}_j}. 
\label{eq:SI_density_mode} 
\end{equation} 
All systems considered here are homogeneous, so $\operatorname{tr}(\hat P\hat{\rho}_{\mathbf q})=0$ for $\mathbf q\neq0$, and the static structure factor is the equal-time density correlation 
\begin{equation} 
S(\mathbf q) = \frac{1}{N} \operatorname{tr} \left( \hat P\hat{\rho}_{\mathbf q}\hat{\rho}_{-\mathbf q} \right), \qquad \mathbf q\neq0. 
\label{eq:SI_static_structure} 
\end{equation} 

Density fluctuations may also be correlated at different times, in real time or in imaginary time. In real time, the corresponding equilibrium correlation function is 
\begin{equation} 
f(\mathbf q,t) = \frac{1}{N} \operatorname{tr} \left[ \hat P \hat{\rho}_{\mathbf q}(t) \hat{\rho}_{-\mathbf q}(0) \right], 
\label{eq:SI_real_time_density} 
\end{equation} 
where 
\begin{equation} 
\hat A(t) = e^{i\hat{\mathcal H}t/\hbar} \hat A e^{-i\hat{\mathcal H}t/\hbar}. 
\end{equation} 
Its real-frequency Fourier transform is the dynamic structure factor (\cref{SI:spectral}). At $t=0$, \cref{eq:SI_real_time_density} reduces to the static structure factor, $f(\mathbf q,0)=S(\mathbf q)$. The equilibrium counterpart relevant to the imaginary-time formulation is obtained by the Wick rotation $t\to -i\tau$; we keep the symbol $f$ for the resulting function of $\tau$,
\begin{equation} 
f(\mathbf q,\tau) = \frac{1}{N} \operatorname{tr} \left[ \hat P \hat{\rho}_{\mathbf q}^{(\tau)} \hat{\rho}_{-\mathbf q}^{(0)} \right], \qquad 0\leq\tau\leq\beta\hbar, 
\label{eq:SI_imaginary_time_density} 
\end{equation} 
where 
\begin{equation} 
\hat A^{(\tau)} = e^{\tau\hat{\mathcal H}/\hbar} \hat A e^{-\tau\hat{\mathcal H}/\hbar} 
\label{eq:SI_imaginary_time_evolution} 
\end{equation} 
denotes imaginary-time evolution. Although $f(\mathbf q,\tau)$ contains information about the excitation spectrum, it is an equilibrium correlation function and does not describe real-time dynamics.

Two properties of $f(\mathbf q,\tau)$ are used throughout: it is real and positive, and it is symmetric about $\tau=\beta\hbar/2$. In the eigenbasis of the Hamiltonian, $\hat{\mathcal H}|m\rangle=E_m|m\rangle$, with thermal weights $p_m=e^{-\beta E_m}/Z$, it is a sum of positive terms~\cite{dornheim2023physical},
\begin{equation}
f(\mathbf q,\tau)=\frac{1}{N}\sum_{m,l}p_m\,e^{\tau(E_m-E_l)/\hbar}\,\big|\langle m|\hat\rho_{\mathbf q}|l\rangle\big|^2 .
\label{eq:SI_f_lehmann}
\end{equation}
We assume that the Hamiltonian is invariant under time reversal (no magnetic field), so that the path-integral action of \cref{SI:path_integral} is real and $f(-\mathbf q,\tau)=f(\mathbf q,\tau)$. The Kubo--Martin--Schwinger relation, $f(\mathbf q,\beta\hbar-\tau)=f(-\mathbf q,\tau)$, then gives~\cite{dornheim2023physical,tolias2024fourier,dornheim2024dynamic}
\begin{equation}
f(\mathbf q,\beta\hbar-\tau)=f(\mathbf q,\tau) .
\label{eq:SI_f_symmetric}
\end{equation}
In particular $f(\mathbf q,\beta\hbar)=f(\mathbf q,0)=S(\mathbf q)$, so the continuation of $f(\mathbf q,\tau)$ with period $\beta\hbar$ is continuous.

The (Matsubara) density-density correlation function, whose sum over frequencies is the static structure factor, is the Fourier transform of \cref{eq:SI_imaginary_time_density} along the imaginary-time interval, 
\begin{equation} 
S(\mathbf q,\omega_n) = \frac{1}{\hbar} \int_0^{\beta\hbar} \mathrm d\tau\, e^{i\omega_n\tau} f(\mathbf q,\tau) = \frac{1}{N\hbar} \int_0^{\beta\hbar} \mathrm d\tau\, e^{i\omega_n\tau} \operatorname{tr} \left[ \hat P \hat{\rho}_{\mathbf q}^{(\tau)} \hat{\rho}_{-\mathbf q}^{(0)} \right], 
\label{eq:SI_Matsubara_transform} 
\end{equation} 
where 
\begin{equation} 
\omega_n = \frac{2\pi n}{\beta\hbar}, \qquad n\in\mathbb Z, 
\end{equation} 
are bosonic Matsubara frequencies. Because $f(\mathbf q,\tau)$ is real and symmetric about $\tau=\beta\hbar/2$, $S(\mathbf q,\omega_n)$ is real and even in $n$, and the inverse transform is the Fourier--Matsubara series~\cite{tolias2024fourier} 
\begin{equation} 
f(\mathbf q,\tau) = \frac{1}{\beta} \sum_{n=-\infty}^{\infty} e^{-i\omega_n\tau} S(\mathbf q,\omega_n). 
\label{eq:SI_inverse_Matsubara} 
\end{equation} 
Evaluating it at $\tau=0$ recovers the static structure factor, 
\begin{equation} 
S(\mathbf q) = \frac{1}{\beta} \sum_{n=-\infty}^{\infty} S(\mathbf q,\omega_n). 
\label{eq:SI_static_from_Matsubara} 
\end{equation} 
The static structure factor therefore contains only the sum over Matsubara components, whereas the complete set $\{S(\mathbf q,\omega_n)\}_n$ resolves how the density correlations are distributed along imaginary time.

Equivalently, $S(\mathbf q,\omega_n)$ is the covariance of the Fourier--Matsubara density modes evaluated on the imaginary-time paths. It is obtained by Fourier transforming $\hat{\rho}_{\mathbf q}$ of \cref{eq:SI_density_mode} along the imaginary-time direction. In the path-integral representation, the density mode evaluated on a set of paths $R=\{\mathbf r_j^{(\tau)}\}$ is 
\begin{equation} 
\rho_{\mathbf q,n}[R] = \frac{1}{\sqrt{\beta}\hbar} \int_0^{\beta\hbar} \mathrm d\tau\, e^{i\omega_n\tau} \sum_{j=1}^{N} e^{-i\mathbf q\cdot\mathbf r_j^{(\tau)}} \equiv \frac{1}{\sqrt{\beta}\hbar} \int_0^{\beta\hbar} \mathrm d\tau\, e^{i\omega_n\tau} \rho_{\mathbf q}^{(\tau)} . 
\label{eq:SI_path_density_Matsubara_mode} 
\end{equation} 
Since path averages of products at different imaginary times are imaginary-time-ordered correlation functions, \cref{eq:SI_imaginary_time_density} reads, in the path form,
\begin{equation}
\big\langle\rho_{\mathbf q}^{(\tau)}\rho_{-\mathbf q}^{(\tau')}\big\rangle=Nf(\mathbf q,\tau-\tau') .
\label{eq:SI_path_two_point}
\end{equation}
Inserting \cref{eq:SI_path_density_Matsubara_mode} for both modes,
\begin{equation}
\big\langle\rho_{\mathbf q,n}[R]\,\rho_{-\mathbf q,-n}[R]\big\rangle=\frac{N}{\beta\hbar^2}\int_0^{\beta\hbar}\mathrm d\tau\int_0^{\beta\hbar}\mathrm d\tau'\,e^{i\omega_n(\tau-\tau')}f(\mathbf q,\tau-\tau')=\frac{N}{\hbar}\int_0^{\beta\hbar}\mathrm du\,e^{i\omega_nu}f(\mathbf q,u) ,
\end{equation}
since the integrand depends only on $u=\tau-\tau'$ and has period $\beta\hbar$. By \cref{eq:SI_Matsubara_transform} the last integral is $NS(\mathbf q,\omega_n)$, so
\begin{equation} 
S(\mathbf q,\omega_n) = \frac{1}{N} \left\langle \rho_{\mathbf q,n}[R] \rho_{-\mathbf q,-n}[R] \right\rangle = \frac{1}{N} \left\langle |\rho_{\mathbf q,n}[R]|^2 \right\rangle , 
\label{eq:SI_Matsubara_mode_covariance} 
\end{equation} 
since $\rho_{-\mathbf q,-n}[R]=\rho_{\mathbf q,n}[R]^*$ on every path. With operators in place of the paths, \cref{eq:SI_Matsubara_mode_covariance} holds only with explicit imaginary-time ordering. The modes also obey Parseval's identity,
\begin{equation}
    \sum_n\big|\rho_{\mathbf q,n}[R]\big|^2 = \frac{1}{\hbar}\int_0^{\beta\hbar}\mathrm{d}\tau\;\rho_{\mathbf q}^{(\tau)}\rho_{-\mathbf q}^{(\tau)} .
\label{eq:SI_parseval}
\end{equation}

\subsection{Spectral representation and density response}
\label{SI:spectral}

In the eigenbasis of the Hamiltonian, $S(\mathbf q,\omega_n)$ takes a spectral (Lehmann) form. Each term of \cref{eq:SI_f_lehmann} depends on $\tau$ through $e^{-a\tau/\hbar}$, with $a=E_l-E_m$, and since $e^{i\omega_n\beta\hbar}=1$,
\begin{equation}
    \frac{1}{\hbar}\int_0^{\beta\hbar}\mathrm{d}\tau\,e^{i\omega_n\tau}e^{-a\tau/\hbar}
    =\frac{1-e^{-\beta a}}{a-i\hbar\omega_n}\quad(a\neq0),
    \qquad
    \frac{1}{\hbar}\int_0^{\beta\hbar}\mathrm{d}\tau\,e^{i\omega_n\tau}=\beta\,\delta_{n0} .
\label{eq:SI_tau_integral}
\end{equation}
With $p_m\big(1-e^{-\beta(E_l-E_m)}\big)=p_m-p_l$, and averaging over $\pm\omega_n$, since $S(\mathbf q,\omega_n)$ is even in $n$, \cref{eq:SI_Matsubara_transform} gives
\begin{equation}
    S(\mathbf q,\omega_n)=\frac{1}{N}\sum_{\substack{m,l\\E_l\neq E_m}}|\langle m|\hat{\rho}_{\mathbf q}|l\rangle|^2\,\frac{(p_m-p_l)(E_l-E_m)}{(\hbar\omega_n)^2+(E_l-E_m)^2}
    +\delta_{n0}\,\frac{\beta}{N}\sum_{\substack{m,l\\E_l=E_m}}p_m\,|\langle m|\hat{\rho}_{\mathbf q}|l\rangle|^2 ,
\label{eq:SI_spectral}
\end{equation}
where the degenerate pairs, $E_l=E_m$, contribute only at $n=0$. Since $p_m>p_l$ exactly when $E_l>E_m$, every term is non-negative: $S(\mathbf q,\omega_n)$ is positive and decreases monotonically with $|n|$.

Both $f(\mathbf q,\tau)$ and $S(\mathbf q,\omega_n)$ follow from the dynamic structure factor measured in scattering experiments, the real-frequency transform of \cref{eq:SI_real_time_density},
\begin{equation}
S(\mathbf q,\omega)=\frac{1}{2\pi N}\int_{-\infty}^{\infty}\mathrm dt\,e^{i\omega t}\operatorname{tr}\big[\hat P\hat\rho_{\mathbf q}(t)\hat\rho_{-\mathbf q}(0)\big]=\frac{1}{N}\sum_{m,l}p_m\big|\langle m|\hat\rho_{\mathbf q}|l\rangle\big|^2\,\delta\big(\omega-(E_l-E_m)/\hbar\big) ,
\label{eq:SI_dsf}
\end{equation}
Comparison with \cref{eq:SI_f_lehmann} gives $f(\mathbf q,\tau)=\int\mathrm d\omega\,S(\mathbf q,\omega)\,e^{-\omega\tau}$, the two-sided Laplace transform through which $f(\mathbf q,\tau)$ is obtained from measured spectra~\cite{dornheim2022accurate}, and with \cref{eq:SI_Matsubara_transform},
\begin{equation}
S(\mathbf q,\omega_n)=\frac{1}{\hbar}\int_{-\infty}^{\infty}\mathrm d\omega\,S(\mathbf q,\omega)\,\frac{1-e^{-\beta\hbar\omega}}{\omega-i\omega_n}=\frac{2}{\hbar}\int_0^{\infty}\mathrm d\omega\,S(\mathbf q,\omega)\big(1-e^{-\beta\hbar\omega}\big)\frac{\omega}{\omega^2+\omega_n^2} ,
\label{eq:SI_Matsubara_from_dsf}
\end{equation}
where the second form uses detailed balance, $S(\mathbf q,-\omega)=e^{-\beta\hbar\omega}S(\mathbf q,\omega)$~\cite{dornheim2023physical}.

The density response function in imaginary time~\cite{mahan2013many,giuliani2005quantum,tolias2024fourier}, used in \cref{sec:rpa}, is
\begin{equation}
\chi(\mathbf q,i\omega_n)=-\bar\rho\,S(\mathbf q,\omega_n) ,
\label{eq:SI_chi_S}
\end{equation}
and for the ideal gas $\chi_0=-\bar\rho S_0$. At $n=0$ it is the static isothermal response~\cite{dornheim2023physical,tolias2024fourier}, and for $n\neq0$ it equals the retarded response $\chi(\mathbf q,\omega)$ at $\omega=i|\omega_n|$~\cite{tolias2024fourier}.

\subsection{High-frequency tail and the classical limit}
\label{SI:high_frequency}

For $n\neq0$ the degenerate term of \cref{eq:SI_spectral} is absent, and expanding each remaining term in powers of $(\hbar\omega_n)^{-2}$ gives
\begin{equation}
    S(\mathbf q,\omega_n)=\frac{M_1(\mathbf q)}{(\hbar\omega_n)^2}+O\big(\omega_n^{-4}\big),\qquad
    M_1(\mathbf q)\equiv\frac{1}{N}\sum_{m,l}(p_m-p_l)(E_l-E_m)\,|\langle m|\hat{\rho}_{\mathbf q}|l\rangle|^2 .
\label{eq:SI_tail}
\end{equation}
On a lattice the spectrum is bounded, and the expansion converges for large $|n|$; in the continuum it is asymptotic. $M_1$ is a thermal average of a double commutator~\cite{dornheim2023physical},
\begin{equation}
    M_1(\mathbf q)=\frac{1}{N}\big\langle\big[\hat{\rho}_{\mathbf q},[\hat{\mathcal{H}},\hat{\rho}_{-\mathbf q}]\big]\big\rangle .
\label{eq:SI_highfreq}
\end{equation}
Since every term of \cref{eq:SI_spectral} is non-negative and $(\hbar\omega_n)^2+(E_l-E_m)^2\ge(\hbar\omega_n)^2$, the leading term of \cref{eq:SI_tail} bounds $S(\mathbf q,\omega_n)$ for any spectrum,
\begin{equation}
S(\mathbf q,\omega_n)\le\frac{M_1(\mathbf q)}{(\hbar\omega_n)^2},\qquad n\neq0 ,
\label{eq:SI_bound}
\end{equation}
so the sum in \cref{eq:SI_static_from_Matsubara} converges.

For $\hat{\mathcal{H}}=\sum_j\hat{\mathbf p}_j^2/2m+U(\hat{\mathbf r}_1,\dots,\hat{\mathbf r}_N)$ the potential commutes with $\hat{\rho}_{-\mathbf q}$, and the double commutator is the constant $2N\epsilon_q$. \Cref{eq:SI_tail,eq:SI_highfreq} then give the f-sum rule~\cite{dornheim2023physical,dornheim2024dynamic},
\begin{equation}
    M_1(\mathbf q)=2\epsilon_q,\qquad
    S(\mathbf q,\omega_n)=\frac{2\epsilon_q}{(\hbar\omega_n)^2}+O(\omega_n^{-4})=\frac{q^2}{m\omega_n^2}+O(\omega_n^{-4}),
\label{eq:SI_fsum}
\end{equation}
independent of the interaction and of the statistics.

\emph{Classical limit.} Consider $\hbar\to0$ at fixed $\beta$, $m$, $\mathbf q$ and interactions, so that $\hbar\omega_n=2\pi n/\beta$ stays fixed. The transition energies are $E_l-E_m=\hbar\nu$, where the frequencies $\nu$ that carry the spectral weight stay finite: they are those of the classical dynamic structure factor~\cite{hansen2006theory}. For $n\neq0$, $\omega_n\to\infty$, so $(E_l-E_m)^2/(\hbar\omega_n)^2=\nu^2/\omega_n^2\to0$ in \cref{eq:SI_spectral}, and only the leading term of \cref{eq:SI_tail} remains. By \cref{eq:SI_fsum}, $S(\mathbf q,\omega_n)\to2\epsilon_q/(\hbar\omega_n)^2$ for $S$ and $S_0$ alike, so $S(\mathbf q,\omega_n)/S_0(\mathbf q,\omega_n)\to1$ for every $n\neq0$.

For $n=0$, differentiating \cref{eq:SI_f_lehmann} at $\tau=0$ and using $f(-\mathbf q,\tau)=f(\mathbf q,\tau)$ shows that $f(\mathbf q,\tau)$ starts at $S(\mathbf q)$ with slope $-M_1/2\hbar=-\epsilon_q/\hbar$. Each term of \cref{eq:SI_f_lehmann} is convex in $\tau$, and $f$ is symmetric about $\tau=\beta\hbar/2$ [\cref{eq:SI_f_symmetric}], so $f$ stays within $(\epsilon_q/\hbar)(\beta\hbar/2)=\beta\epsilon_q/2$ of $S(\mathbf q)$. By \cref{eq:SI_Matsubara_transform}, $S(\mathbf q,0)$ is then within $\beta^2\epsilon_q/2$ of $\beta S(\mathbf q)$. Since $\epsilon_q\propto\hbar^2$, $S(\mathbf q,0)\to\beta S(\mathbf q)$ and, for the ideal gas, $S_0(\mathbf q,0)\to\beta$.

\section{Derivation of the closed form}

\subsection{Path-integral representation}
\label{SI:path_integral}

Here we briefly recap how to evaluate the trace in the position representation, since we use it in the main steps of our derivation. Consider $\operatorname{tr}e^{-\beta\hat{\mathcal H}}$ with $\hat{\mathcal H}=\hat{\mathcal H}_0+\hat U$, $\hat{\mathcal H}_0=\sum_j\hat{\mathbf p}_j^2/2m$ and $\hat U=U(\hat{\mathbf R})$, where $\mathbf R=(\mathbf r_1,\dots,\mathbf r_N)$. In the position representation it is~\cite{altland2010condensed}
\begin{equation}
\operatorname{tr}e^{-\beta(\hat{\mathcal H}_0+\hat U)}=\oint\mathcal D\mathbf R\,\exp\Big[-\mathcal A_0[R]-\frac1\hbar\int_0^{\beta\hbar}\mathrm d\tau\,U\big(\mathbf R^{(\tau)}\big)\Big] ,\qquad
\mathcal{A}_0[R] \equiv \frac{1}{\hbar}\int_0^{\beta\hbar}\mathrm{d}\tau \sum_{j=1}^N\frac{1}{2}m\left|\dot{\mathbf{r}}_j^{(\tau)}\right|^2 ,
\label{eq:SI_path_integral_boltzmann}
\end{equation}
where $\oint\mathcal D\mathbf R$, the path measure of \cref{eq:path_measure}, runs over closed paths in imaginary time, and $\mathcal A_0$ is the free-particle Euclidean action. The mass enters only through $e^{-\mathcal A_0[R]}$ and the normalization of the measure, which is fixed by the free-particle propagator and does not depend on the paths.

For indistinguishable particles, quantum statistics enter only through the boundary condition of the paths, which close up to a permutation of the particles. The trace over symmetric (bosons) or antisymmetric (fermions) states is written with the projector $(N!)^{-1}\sum_\sigma\eta^{P_\sigma}\hat\Pi_\sigma$, where $\hat\Pi_\sigma$ permutes the particle labels according to $\sigma$. The projector acts only at the end of the imaginary-time interval, where it connects $\mathbf r_j^{(\beta\hbar)}$ to $\mathbf r_{\sigma(j)}^{(0)}$, so some particles do not return to their own starting points; this is the sum over permutations in \cref{eq:path_measure}. For fermions the reference path weight, $P_0[R]=e^{-\mathcal A_0[R]}/\mathcal I[0]$, carries the sign $\eta^{P_\sigma}$.

\subsection{Maximum relative-entropy inference}
\label{SI:maximization}

The classical derivation~\cite{ariel2020inferring} maximized the entropy. However, it is not the correct object for quantum systems: maximizing the von Neumann entropy $H=-\operatorname{tr}(\hat P\ln\hat P)$ subject only to the density constraints and $\operatorname{tr}\hat P=1$ would assign a flat prior to $\hat P$ and ignore the kinetic energy, which every state of the system carries. We therefore maximize the negative quantum relative entropy~\cite{vedral2002role}, \cref{eq:H_rel_def}, with respect to the thermal state of the non-interacting gas, $\hat P_0=e^{-\beta\hat{\mathcal H}_0}/\mathcal I[0]$, where $\hat{\mathcal H}_0$ is the kinetic-energy operator and the trace runs over symmetric (antisymmetric) states for bosons (fermions). $H_{\text{rel}}=-D(\hat P\|\hat P_0)$ is minus the quantum Kullback--Leibler divergence, so $H_{\text{rel}}\le0$, with equality only for $\hat P=\hat P_0$ (Klein's inequality). Using $\ln\hat P_0=-\beta\hat{\mathcal H}_0-\ln\mathcal I[0]$ and $\operatorname{tr}\hat P=1$,
\begin{equation}
H_{\text{rel}}=H-\beta\operatorname{tr}\big(\hat P\hat{\mathcal H}_0\big)-\ln\mathcal I[0] .
\label{eq:SI_Hrel_free_energy_form}
\end{equation}
Maximizing $H_{\text{rel}}$ is therefore maximizing $H$ with the kinetic energy weighted by the physical $\beta$.

The relative entropy differs from the excess entropy by the interaction-induced change in kinetic energy. Applying \cref{eq:SI_Hrel_free_energy_form} to $\hat P_0$, whose entropy is $H_0=\beta E_{\text{kin}}^{0}+\ln\mathcal I[0]$ with $E_{\text{kin}}^{0}=\operatorname{tr}(\hat P_0\hat{\mathcal H}_0)$, relates $H_{\text{rel}}$ to the excess entropy $H_{\text{ex}}\equiv H-H_0$,
\begin{equation}
H_{\text{rel}}=H_{\text{ex}}+\beta\big(E_{\text{kin}}^{0}-E_{\text{kin}}\big),\qquad E_{\text{kin}}=\operatorname{tr}\big(\hat P\hat{\mathcal H}_0\big) ,
\label{eq:SI_Hrel_Hex}
\end{equation}
which is the relation stated in \cref{sec:legendre} of the End Matter.

We now carry out the maximization in the position representation, over the statistical weights $P[R]$ of the paths. The relative entropy of the path weights with respect to the reference weight $P_0[R]$ is
\begin{equation}
    H_{\text{rel}}[P] = -\oint\mathcal{D}\mathbf{R}\,P[R]\ln\frac{P[R]}{P_0[R]} .
\label{eq:SI_Hrel_paths}
\end{equation}
With the multipliers $\lambda_{\mathbf q,n}$, and $\mu$ for the normalization, the Lagrangian is
\begin{equation}
    \mathcal{L} = H_{\text{rel}}[P] - \sum_{\mathbf{q},n}\lambda_{\mathbf{q},n}\left[\oint\mathcal{D}\mathbf{R}\,P[R]\,|\rho_{\mathbf{q},n}[R]|^2 - NS(\mathbf{q},\omega_n)\right] - \mu\left(\oint\mathcal{D}\mathbf{R}\,P[R]-1\right) .
\end{equation}
Setting $\delta\mathcal L/\delta P[R]=0$ and normalizing gives
\begin{equation}
    P[R] = \mathcal{Z}^{-1}\exp\left(-\mathcal{A}_0[R]-\sum_{\mathbf{q},n}\lambda_{\mathbf{q},n}\left[|\rho_{\mathbf{q},n}[R]|^2-NS(\mathbf{q},\omega_n)\right]\right),
\label{eq:effectiveZ}
\end{equation}
with the effective partition function
\begin{equation}
    \mathcal{Z} = \oint\mathcal{D}\mathbf{R}\, \exp \left(-\mathcal{A}_0[R] -\sum_{\mathbf{q},n}\lambda_{\mathbf{q},n}\left[ |{\rho}_{\mathbf{q},n}[R]|^2 -NS(\mathbf{q},\omega_n)\right]\right) .
\label{eq:effectiveZ_PI}
\end{equation}

The stationary point is the maximum of $H_{\text{rel}}[P]$ for bosons and distinguishable particles, whose path weights are positive, and a saddle point for fermions. Since $\delta\mathcal L/\delta P[R]=0$ and the constraints are linear in $P$, the first-order change of $H_{\text{rel}}$ vanishes for any $\delta P$ that preserves the normalization and the constraints. The second-order change follows from the second derivative of $-P\ln(P/P_0)$, which is $-1/P$:
\begin{equation}
H_{\text{rel}}[P+\delta P]=H_{\text{rel}}[P]-\oint\mathcal D\mathbf R\,\frac{(\delta P[R])^2}{2P[R]}+O(\delta P^3) .
\label{eq:SI_second_variation}
\end{equation}
For bosons and distinguishable particles the path weights are probabilities, so $H_{\text{rel}}[P]$ is strictly concave, and on the convex set of probabilities that satisfy the linear constraints it has at most one stationary point, its global maximum. For fermions the paths with odd permutations have negative weight [\cref{eq:path_measure}]. The ratio $P[R]/P_0[R]$ is still positive, so $H_{\text{rel}}[P]$ and its stationarity conditions remain defined, but the second-order term is positive for variations confined to odd permutations and negative for even ones: the stationary point is a saddle point, and for fermions the construction is defined by stationarity.

Dividing \cref{eq:effectiveZ} by $P_0[R]$ and taking the logarithm gives
\begin{equation}
\ln\frac{P[R]}{P_0[R]}=-\sum_{\mathbf q,n}\lambda_{\mathbf q,n}\big[|\rho_{\mathbf q,n}[R]|^2-NS(\mathbf q,\omega_n)\big]-\ln\mathcal Z+\ln\mathcal I[0] .
\end{equation}
Averaged over $P[R]$, the bracket vanishes when the constraints hold, and \cref{eq:SI_Hrel_paths} gives
\begin{equation}
    H_{\text{rel}}=\ln\mathcal{Z} - \ln \mathcal{I}[0] .
\label{eq:SI_Hrel_lnZ}
\end{equation}
Since
\begin{equation}
\frac{\partial\ln\mathcal Z}{\partial\lambda_{\mathbf q,n}}=NS(\mathbf q,\omega_n)-\big\langle|\rho_{\mathbf q,n}[R]|^2\big\rangle_P ,
\label{eq:SI_dlnZ}
\end{equation}
the constraints are the stationarity conditions of $\mathcal Z$ with respect to the multipliers, so we can extremize $\mathcal Z$ instead, which we do from \cref{SI:field} on.

\subsection{Collective variables and the conjugate field}
\label{SI:field}

We now evaluate $\mathcal Z$ of \cref{eq:effectiveZ_PI} for multipliers that may depend on $n$, following the construction of Ariel and Diamant [Appendix~A of \cite{ariel2020inferring}]: collective variables for the density modes are introduced through delta functions, which are written with a conjugate field. We omit constants that do not depend on the multipliers; they are fixed in \cref{SI:closed_form}.

We introduce collective variables $\tilde\rho_{\mathbf q,n}$, with $\tilde\rho_{-\mathbf q,-n}=\tilde\rho_{\mathbf q,n}^*$, through $1=\int\mathcal D\tilde\rho\;\delta\big[\tilde\rho-\rho[R]\big]$, and a conjugate field $\tilde\psi_{\mathbf q,n}$, with $\tilde\psi_{-\mathbf q,-n}=\tilde\psi_{\mathbf q,n}^*$,
\begin{equation}
\delta\big[\tilde\rho-\rho[R]\big]\propto\int\mathcal D\tilde\psi\;\exp\Big[i\sum_{\mathbf q,n}\tilde\psi_{\mathbf q,n}\big(\tilde\rho_{\mathbf q,n}-\rho_{\mathbf q,n}[R]\big)\Big] .
\label{eq:SI_psi_def}
\end{equation}
Under the delta functions $|\rho_{\mathbf q,n}[R]|^2$ may be replaced by $|\tilde\rho_{\mathbf q,n}|^2$, and inserting \cref{eq:SI_psi_def} into \cref{eq:effectiveZ_PI} gives
\begin{equation}
\mathcal Z\propto e^{N\sum_{\mathbf q,n}\lambda_{\mathbf q,n}S(\mathbf q,\omega_n)}\int\mathcal D\tilde\rho\,\mathcal D\tilde\psi\;\mathcal I[\tilde\psi]\,\exp\Big[\sum_{\mathbf q,n}\big(i\tilde\psi_{\mathbf q,n}\tilde\rho_{\mathbf q,n}-\lambda_{\mathbf q,n}|\tilde\rho_{\mathbf q,n}|^2\big)\Big] ,
\label{eq:SI_Z_rho_psi}
\end{equation}
where the paths now appear only in
\begin{equation}
\mathcal{I}[\tilde{\psi}] \equiv \oint\mathcal{D}\mathbf{R}\exp \Big(-\mathcal{A}_0[R] -i \sum_{\mathbf{q},n} \tilde{\psi}_{\mathbf{q},n} {\rho}_{\mathbf{q},n}[R]\Big) .
\label{eq:SI_I_psi}
\end{equation}
For indistinguishable particles the permutations in $\oint\mathcal D\mathbf R$ couple the particles, so $\mathcal I[\tilde\psi]$ does not factorize over particles. Ariel and Diamant, treating distinguishable classical particles, factorized it; the derivation below does not require this.

Integrating \cref{eq:SI_Z_rho_psi} over $\tilde\rho_{\mathbf q,n}$ gives
\begin{equation}
\mathcal Z\propto e^{N\sum_{\mathbf q,n}\lambda_{\mathbf q,n}S(\mathbf q,\omega_n)}\Big(\prod_{\mathbf q,n}\lambda_{\mathbf q,n}^{-1/2}\Big)\,Z_\psi ,
\label{eq:effective_Z_almost_done}
\end{equation}
with
\begin{equation}
Z_{\psi} = \int\mathcal D\tilde\psi\; e^{-\mathcal{H}_{\psi}[\tilde\psi]}, \qquad \mathcal{H}_{\psi} = \sum_{\mathbf{q},n}   \frac{|\tilde{\psi}_{\mathbf{q},n}|^2}{4\lambda_{\mathbf{q},n}} - \ln \mathcal{I}[\tilde{\psi}] ,
\label{eq:Z_psi}
\end{equation}
which is \cref{eq:Zaux} of the End Matter.

For $\lambda_{\mathbf q,n}<0$ the Gaussian integral over $\tilde\rho_{\mathbf q,n}$ diverges on the real axis, and \cref{eq:effective_Z_almost_done} is defined by analytic continuation from $\lambda_{\mathbf q,n}>0$. Negative multipliers occur in two of the tests of the main text, the ring and the electron gas. The continuation is unambiguous because $\mathcal Z$ itself is analytic in the multipliers: the modes are bounded on every path, $|\rho_{\mathbf q,n}[R]|\le N\sqrt\beta$ by \cref{eq:SI_path_density_Matsubara_mode} and $\sum_n|\rho_{\mathbf q,n}[R]|^2\le\beta N^2$ by \cref{eq:SI_parseval}, so for a finite set of modes the integrand of \cref{eq:effectiveZ_PI} is an entire function of the multipliers, bounded uniformly in $R$ on bounded sets of them. The Gaussian approximation (\cref{SI:Approximation}) is continued in the same way.

Before we continue to the Gaussian approximation, we provide below a physical interpretation of $\mathcal I[\tilde\psi]$. This functional describes the ideal quantum gas in an external field. With \cref{eq:SI_path_density_Matsubara_mode},
\begin{equation}
\sum_{\mathbf q,n}\tilde\psi_{\mathbf q,n}\rho_{\mathbf q,n}[R]=\frac1\hbar\int_0^{\beta\hbar}\mathrm d\tau\sum_{j=1}^N\tilde\psi\big(\mathbf r_j^{(\tau)},\tau\big),\qquad\tilde\psi(\mathbf r,\tau)\equiv\frac{1}{\sqrt\beta}\sum_{\mathbf q,n}\tilde\psi_{\mathbf q,n}\,e^{i(\omega_n\tau-\mathbf q\cdot\mathbf r)} ,
\end{equation}
so that
\begin{equation}
\mathcal I[\tilde\psi]=\oint\mathcal D\mathbf R\,\exp\Big[-\frac1\hbar\int_0^{\beta\hbar}\mathrm d\tau\sum_{j=1}^N\Big(\frac12m\big|\dot{\mathbf r}_j^{(\tau)}\big|^2+i\tilde\psi\big(\mathbf r_j^{(\tau)},\tau\big)\Big)\Big] .
\end{equation}
The field $\tilde\psi(\mathbf r,\tau)$ is real because $\tilde\psi_{-\mathbf q,-n}=\tilde\psi_{\mathbf q,n}^*$. Thus $\mathcal I[\tilde\psi]$ is the partition function of the ideal quantum gas in the imaginary external potential $i\tilde\psi(\mathbf r,\tau)$, which depends on imaginary time.

\subsection{The Gaussian approximation}
\label{SI:Approximation}

So far the derivation steps are exact; here we show the single approximation that gives the closed-form expressions. We expand $\ln\mathcal I[\tilde\psi]$ in powers of the conjugate field,
\begin{equation}
\ln\mathcal I[\tilde\psi]=\ln\mathcal I[0]+\sum_{\mathbf q,n}\frac{\delta\ln\mathcal I}{\delta\tilde\psi_{\mathbf q,n}}\bigg|_{\tilde\psi=0}\tilde\psi_{\mathbf q,n}+\frac12\sum_{\mathbf q,n}\sum_{\mathbf k,l}\frac{\delta^2\ln\mathcal I}{\delta\tilde\psi_{\mathbf q,n}\,\delta\tilde\psi_{\mathbf k,l}}\bigg|_{\tilde\psi=0}\tilde\psi_{\mathbf q,n}\tilde\psi_{\mathbf k,l}+O(\tilde\psi^3) .
\label{eq:SI_cumulant_expansion}
\end{equation}
The Gaussian approximation of the classical derivation~\cite{ariel2020inferring} keeps the terms up to second order; the neglected terms are set by the three- and higher-point connected density correlations of the reference.

The first-order term vanishes. By \cref{eq:SI_I_psi}, the first derivative in \cref{eq:SI_cumulant_expansion} is $-i\langle\rho_{\mathbf q,n}\rangle_0$, with $\langle\cdot\rangle_0$ the average in the reference weight $P_0[R]$. This weight is invariant under a common translation of the paths, $\mathbf r_j^{(\tau)}\to\mathbf r_j^{(\tau)}+\mathbf a$, and under a common shift in imaginary time, $\mathbf r_j^{(\tau)}\to\mathbf r_j^{(\tau+\tau_0)}$, under which the modes of \cref{eq:SI_path_density_Matsubara_mode} transform as
\begin{equation}
\rho_{\mathbf q,n}[R]\to e^{-i\mathbf q\cdot\mathbf a}\rho_{\mathbf q,n}[R],\qquad\rho_{\mathbf q,n}[R]\to e^{-i\omega_n\tau_0}\rho_{\mathbf q,n}[R] .
\label{eq:SI_mode_symmetries}
\end{equation}
Hence $\langle\rho_{\mathbf q,n}\rangle_0=e^{-i\mathbf q\cdot\mathbf a}\langle\rho_{\mathbf q,n}\rangle_0$ for every $\mathbf a$, so $\langle\rho_{\mathbf q,n}\rangle_0=0$ for $\mathbf q\neq0$.

By \cref{eq:SI_I_psi}, the second derivative in \cref{eq:SI_cumulant_expansion} is $-\langle\rho_{\mathbf q,n}\rho_{\mathbf k,l}\rangle_0+\langle\rho_{\mathbf q,n}\rangle_0\langle\rho_{\mathbf k,l}\rangle_0=-\langle\rho_{\mathbf q,n}\rho_{\mathbf k,l}\rangle_0$, and the second-order term is $-\frac12\sum_{\mathbf q,n}\sum_{\mathbf k,l}\langle\rho_{\mathbf q,n}\rho_{\mathbf k,l}\rangle_0\tilde\psi_{\mathbf q,n}\tilde\psi_{\mathbf k,l}$. Both invariances of \cref{eq:SI_mode_symmetries} apply to the product,
\begin{equation}
\langle\rho_{\mathbf q,n}\rho_{\mathbf k,l}\rangle_0=e^{-i(\mathbf q+\mathbf k)\cdot\mathbf a}\,e^{-i(\omega_n+\omega_l)\tau_0}\,\langle\rho_{\mathbf q,n}\rho_{\mathbf k,l}\rangle_0
\end{equation}
for every $\mathbf a$ and $\tau_0$, so the correlation vanishes unless $\mathbf k=-\mathbf q$ and $l=-n$. For $(\mathbf k,l)=(-\mathbf q,-n)$, \cref{eq:SI_Matsubara_mode_covariance} applied to the reference gives $\langle\rho_{\mathbf q,n}\rho_{-\mathbf q,-n}\rangle_0=NS_0(\mathbf q,\omega_n)$, where $S_0$ is the density-density correlation function of the ideal quantum gas (\cref{sec:reference}). With these derivatives, \cref{eq:SI_cumulant_expansion} becomes
\begin{equation}
\ln\mathcal I[\tilde\psi]\approx\ln\mathcal I[0]-\frac N2\sum_{\mathbf q,n}S_0(\mathbf q,\omega_n)\,\big|\tilde\psi_{\mathbf q,n}\big|^2 ,
\label{eq:SI_secondorder}
\end{equation}
which is \cref{eq:secondorder} of the End Matter.

Inserting \cref{eq:SI_secondorder} into \cref{eq:Z_psi},
\begin{equation}
\mathcal H_\psi\approx-\ln\mathcal I[0]+\sum_{\mathbf q,n}\frac{|\tilde\psi_{\mathbf q,n}|^2}{4\Lambda_{\mathbf q,n}},\qquad \Lambda_{\mathbf q,n}\equiv\Big(\frac{1}{\lambda_{\mathbf q,n}}+2NS_0(\mathbf q,\omega_n)\Big)^{-1}=\frac{\lambda_{\mathbf q,n}}{1+2N\lambda_{\mathbf q,n}S_0(\mathbf q,\omega_n)} .
\end{equation}
The Gaussian integral over the field then gives
\begin{equation}
Z_\psi\propto\mathcal I[0]\prod_{\mathbf q,n}\Lambda_{\mathbf q,n}^{1/2} .
\label{eq:Z_psi_approx}
\end{equation}
\Cref{eq:Z_psi_approx} holds for every $\lambda_{\mathbf q,n}$ with $1+2N\lambda_{\mathbf q,n}S_0>0$.

\subsection{The closed form}
\label{SI:closed_form}

Inserting \cref{eq:Z_psi_approx} into \cref{eq:effective_Z_almost_done}, with $\Lambda_{\mathbf q,n}/\lambda_{\mathbf q,n}=(1+2N\lambda_{\mathbf q,n}S_0)^{-1}$,
\begin{equation}
\mathcal Z\approx e^{N\sum_{\mathbf q,n}\lambda_{\mathbf q,n}S(\mathbf q,\omega_n)}\,\mathcal I[0]\prod_{\mathbf q,n}\big(1+2N\lambda_{\mathbf q,n}S_0(\mathbf q,\omega_n)\big)^{-1/2} ,
\end{equation}
where the omitted constant is 1: as $\lambda\to0$ the weight $e^{-|\tilde\psi|^2/4\lambda}$ confines the field to $\tilde\psi\to0$, where the expansion \cref{eq:SI_cumulant_expansion} is exact, and $\mathcal Z\to\mathcal I[0]$ by \cref{eq:effectiveZ_PI}. The relative entropy \cref{eq:SI_Hrel_lnZ} becomes
\begin{equation}
H_{\text{rel}}(\lambda)=\ln\mathcal Z-\ln\mathcal I[0]\approx\sum_{\mathbf q,n}\Big[N\lambda_{\mathbf q,n}S(\mathbf q,\omega_n)-\frac12\ln\big(1+2N\lambda_{\mathbf q,n}S_0(\mathbf q,\omega_n)\big)\Big] ,
\label{eq:SI_Hrel_lambda}
\end{equation}
which is \cref{eq:Hrel_lambda}. The stationarity condition of \cref{SI:maximization}, $\partial H_{\text{rel}}/\partial\lambda_{\mathbf q,n}=0$, gives
\begin{equation}
S(\mathbf q,\omega_n)=\frac{S_0(\mathbf q,\omega_n)}{1+2N\lambda_{\mathbf q,n}S_0(\mathbf q,\omega_n)} ,
\label{eq:SI_stationarity}
\end{equation}
and therefore
\begin{equation}
\lambda_{\mathbf q,n}=\frac{1}{2N}\left[\frac{1}{S(\mathbf q,\omega_n)}-\frac{1}{S_0(\mathbf q,\omega_n)}\right] ,
\end{equation}
which is \cref{eq:lambda}. Since $\partial^2H_{\text{rel}}/\partial\lambda_{\mathbf q,n}^2=2N^2S_0^2/(1+2N\lambda_{\mathbf q,n}S_0)^2>0$, the stationary point is the minimum of $H_{\text{rel}}(\lambda)$ over the multipliers, as in the exact theory for bosons and distinguishable particles (\cref{SI:exact}). At this point $1+2N\lambda_{\mathbf q,n}S_0=S_0/S>0$ for any positive input, so the closed form is always defined.

Substituting $1+2N\lambda_{\mathbf q,n}S_0=S_0/S$ and $N\lambda_{\mathbf q,n}S=\frac12(1-S/S_0)$ into \cref{eq:SI_Hrel_lambda} gives
\begin{equation}
    H_{\text{rel}} \approx \frac{1}{2}\sum_{\mathbf{q},n}\left[\ln\left(\frac{S(\mathbf{q},\omega_n)}{S_0(\mathbf{q},\omega_n)}\right) -\frac{S(\mathbf{q},\omega_n)}{S_0(\mathbf{q},\omega_n)} +1\right] ,
\end{equation}
which is \cref{eq:H_rel_main}. Each term is non-positive, since $\ln x\le x-1$, and vanishes only for $S=S_0$, so $H_{\text{rel}}\le0$, as for the exact relative entropy.

The Gaussian approximation is exact when the density modes of the reference are Gaussian, since the three- and higher-point connected correlations neglected in \cref{SI:Approximation} then vanish. In general, $H_{\text{rel}}(\lambda)$ in the Gaussian approximation is not guaranteed to lie above the exact one, so the closed form is not a proven bound.

In the classical limit (\cref{SI:high_frequency}), $S(\mathbf q,0)\to\beta S(\mathbf q)$, $S_0(\mathbf q,0)\to\beta$, and $S/S_0\to1$ for $n\neq0$, so $\lambda_{\mathbf q,n\neq0}\to0$ and $\beta\lambda_{\mathbf q,0}\to(2N)^{-1}[1/S(\mathbf q)-1]$, the multiplier of Ariel and Diamant [Eq.~(18) of \cite{ariel2020inferring}] divided by their density of modes.

\subsection{Free energy, kinetic shift and excess entropy}
\label{SI:legendre}

Here we derive the closed forms of $F_{\text{ex}}$, the kinetic shift and $H_{\text{ex}}$.

\emph{Kinetic energy as a mass derivative.} For any multipliers, the mass enters the inferred path weight only through its free part, $e^{-\mathcal A_0[R]}$ together with the normalization of the measure (\cref{SI:path_integral}), since $\rho_{\mathbf q,n}[R]$ depends on the paths alone. With $\mathcal Z_\lambda\equiv\mathcal Z\,e^{-N\sum_{\mathbf q,n}\lambda_{\mathbf q,n}S}$, the path integral of \cref{eq:effectiveZ_PI} without its constant factor, the kinetic energies $E_{\text{kin}}$ and $E_{\text{kin}}^{0}$ of \cref{SI:maximization} are the mass derivatives
\begin{equation}
E_{\text{kin}}=\frac{m}{\beta}\Big(\frac{\partial\ln\mathcal Z_\lambda}{\partial m}\Big)_\lambda,\qquad E_{\text{kin}}^{0}=\frac{m}{\beta}\frac{\partial\ln\mathcal I[0]}{\partial m},
\label{eq:SI_Ekin_mass}
\end{equation}
which are path averages of the primitive kinetic-energy estimator of path-integral simulations. With $F_{\text{ex}}\equiv-\beta^{-1}(\ln\mathcal Z_\lambda-\ln\mathcal I[0])$, subtracting the two derivatives gives \cref{eq:kinetic_shift_mass}, $E_{\text{kin}}-E_{\text{kin}}^{0}=-m(\partial F_{\text{ex}}/\partial m)_\lambda$.

\emph{Excess entropy as a Legendre transform.} For any multipliers, the excess entropy $H_{\text{ex}}$ is the Legendre transform of $F_{\text{ex}}$ at fixed multipliers,
\begin{equation}
H_{\text{ex}}=-\beta F_{\text{ex}}+\beta\big(\partial_\beta(\beta F_{\text{ex}})\big)_\lambda .
\label{eq:SI_Legendre}
\end{equation}
To see this, observe that
\begin{equation}
\Big(\beta\frac{\partial}{\partial\beta}+m\frac{\partial}{\partial m}\Big)_\lambda(\beta F_{\text{ex}})=\sum_{\mathbf q,n}\lambda_{\mathbf q,n}\frac{\partial(\beta F_{\text{ex}})}{\partial\lambda_{\mathbf q,n}}=N\sum_{\mathbf q,n}\lambda_{\mathbf q,n}S .
\label{eq:SI_Z_scaling}
\end{equation}
For the first equality, note that $\beta F_{\text{ex}}$ is a dimensionless function of the energies $\beta^{-1}$, $\lambda_{\mathbf q,n}$, $\hbar^2q^2/m$ and $\hbar^2\bar\rho^{2/d}/m$. It is therefore unchanged when all of them are multiplied by the same factor $s$, that is, under $\beta\to\beta/s$, $m\to m/s$ and $\lambda_{\mathbf q,n}\to s\lambda_{\mathbf q,n}$. Its derivative with respect to $s$ at $s=1$, $\big(-\beta\partial_\beta-m\partial_m\big)_\lambda(\beta F_{\text{ex}})+\sum_{\mathbf q,n}\lambda_{\mathbf q,n}\partial(\beta F_{\text{ex}})/\partial\lambda_{\mathbf q,n}$, therefore vanishes, which is the first equality. The second follows from the constraints, $\partial(\beta F_{\text{ex}})/\partial\lambda_{\mathbf q,n}=\langle|\rho_{\mathbf q,n}|^2\rangle=NS$. Since $H_{\text{rel}}=\ln\mathcal Z-\ln\mathcal I[0]=-\beta F_{\text{ex}}+N\sum_{\mathbf q,n}\lambda_{\mathbf q,n}S$, \cref{eq:SI_Hrel_Hex,eq:kinetic_shift_mass} give $H_{\text{ex}}=-\beta F_{\text{ex}}+N\sum_{\mathbf q,n}\lambda_{\mathbf q,n}S-m(\partial_m(\beta F_{\text{ex}}))_\lambda$, and by \cref{eq:SI_Z_scaling} the last two terms are $\beta(\partial_\beta(\beta F_{\text{ex}}))_\lambda$.

\emph{Closed forms.} Since $\ln\mathcal{Z}_{\lambda}=\ln\mathcal{Z}-N\sum_{\mathbf{q},n}\lambda_{\mathbf{q},n}S$ and $H_{\text{rel}}=\ln\mathcal{Z}-\ln\mathcal{I}[0]$, \cref{eq:SI_Hrel_lambda} gives
\begin{equation}
    \beta F_{\text{ex}} = N\sum_{\mathbf{q},n}\lambda_{\mathbf{q},n}S - H_{\text{rel}} \approx \frac{1}{2}\sum_{\mathbf{q},n}\ln\left(1+2N\lambda_{\mathbf{q},n}S_0\right) .
\label{eq:SI_Fex_lambda}
\end{equation}
With $1+2N\lambda_{\mathbf q,n}S_0=S_0/S$ at the multipliers of \cref{eq:lambda},
\begin{equation}
    \beta F_{\text{ex}}\approx\frac{1}{2}\sum_{\mathbf{q},n}\ln\frac{S_0}{S} ,
\label{eq:SI_Fex}
\end{equation}
which is \cref{eq:Fex}.

At fixed multipliers, the mass and the temperature enter \cref{eq:SI_Fex_lambda} only through $S_0$, so the kinetic shift and the excess entropy follow by the chain rule from
\begin{equation}
    \left(\frac{\partial(\beta F_{\text{ex}})}{\partial S_0}\right)_{\lambda}
    = \frac{1}{2}\,\frac{2N\lambda_{\mathbf{q},n}}{1+2N\lambda_{\mathbf{q},n}S_0}
    = \frac{1}{2}\,\frac{S_0-S}{SS_0}\cdot\frac{S}{S_0}
    = \frac{S_0-S}{2S_0^2} ,
\label{eq:SI_dFdS0}
\end{equation}
where the last steps use $2N\lambda_{\mathbf q,n}=(S_0-S)/SS_0$ and $1+2N\lambda_{\mathbf q,n}S_0=S_0/S$. With \cref{eq:kinetic_shift_mass},
\begin{equation}
    \beta\left(E_{\text{kin}}-E_{\text{kin}}^{0}\right)
    = -m\left(\frac{\partial(\beta F_{\text{ex}})}{\partial m}\right)_{\lambda}
    \approx \frac{1}{2}\sum_{\mathbf{q},n}\frac{S-S_0}{S_0^2}\;m\frac{\partial S_0}{\partial m} ,
\label{eq:SI_shift_explicit}
\end{equation}
which is \cref{eq:kinetic_shift_explicit}, and, with \cref{eq:SI_Legendre},
\begin{equation}
    H_{\text{ex}} \approx \frac{1}{2}\sum_{\mathbf{q},n}\ln\frac{S}{S_0} + \frac{\beta}{2}\sum_{\mathbf{q},n}\frac{S_0-S}{S_0^2}\frac{\partial S_0}{\partial\beta}
    = \frac{1}{2}\sum_{\mathbf{q},n}\left[\ln\frac{S}{S_0}-\beta\,\frac{S-S_0}{S_0^2}\frac{\partial S_0}{\partial\beta}\right] ,
\label{eq:SI_Hex}
\end{equation}
which is \cref{eq:Hex_explicit}.

A scaling relation of the reference trades its mass derivative for a temperature derivative. Under the scaling above, $\beta\to\beta/s$ and $m\to m/s$, $S_0$ becomes $S_0/s$, since it has units of $\beta$. The derivative with respect to $s$ at $s=1$ gives
\begin{equation}
m\frac{\partial S_0}{\partial m}+\beta\frac{\partial S_0}{\partial\beta}=S_0 .
\label{eq:SI_S0_scaling}
\end{equation}
With it, the kinetic shift of \cref{eq:SI_shift_explicit} is $\tfrac12\sum_{\mathbf q,n}(S-S_0)S_0^{-2}(S_0-\beta\,\partial S_0/\partial\beta)$, which is $H_{\text{ex}}-H_{\text{rel}}$ from \cref{eq:SI_Hex,eq:H_rel_main}, so the closed forms satisfy \cref{eq:SI_Hrel_Hex} identically.

\subsection{Multipliers independent of $n$}
\label{SI:n_independent}

With $\lambda_{\mathbf q,n}=\lambda_{\mathbf q}$, Parseval's identity, \cref{eq:SI_parseval}, gives $\sum_{\mathbf q,n}\lambda_{\mathbf q}|\rho_{\mathbf q,n}[R]|^2=\hbar^{-1}\int_0^{\beta\hbar}\mathrm d\tau\sum_{\mathbf q}\lambda_{\mathbf q}\rho_{\mathbf q}^{(\tau)}\rho_{-\mathbf q}^{(\tau)}$. By \cref{eq:SI_path_integral_boltzmann}, $P[R]$ of \cref{eq:effectiveZ} is then the path representation of the thermal state $\hat P_\lambda=\exp[-\beta(\hat{\mathcal H}_0+\sum_{\mathbf q}\lambda_{\mathbf q}\hat\rho_{\mathbf q}\hat\rho_{-\mathbf q})]/\mathcal Z_\lambda$, whose interaction, $2V\lambda_{\mathbf q}$, is an instantaneous pair interaction. If the interaction of the system, $U_{\mathbf q}$, is also instantaneous, $\hat P=\hat P_\lambda$ at $2V\lambda_{\mathbf q}=U_{\mathbf q}$, so the exact multipliers do not depend on $n$. With the same multiplier at every $n$, the constraints enter only through their sum over $n$, $\beta S(\mathbf q)$ [\cref{eq:SI_static_from_Matsubara}]. We therefore use $S(\mathbf q)$ alone as the constraint, and \cref{eq:SI_Hrel_lambda} becomes
\begin{equation}
H_{\text{rel}}(\lambda)\approx\sum_{\mathbf q}\Big[N\beta\lambda_{\mathbf q}S(\mathbf q)-\frac12\sum_n\ln\big(1+2N\lambda_{\mathbf q}S_0(\mathbf q,\omega_n)\big)\Big] ,
\label{eq:SI_Hrel_lambda_static}
\end{equation}
which depends on the interacting system only through $S(\mathbf q)$. The stationarity condition, $\partial H_{\text{rel}}/\partial\lambda_{\mathbf q}=0$, is the sum of \cref{eq:SI_stationarity} over $n$,
\begin{equation}
S(\mathbf q)=\frac{1}{\beta}\sum_n\frac{S_0(\mathbf q,\omega_n)}{1+2N\lambda_{\mathbf q}S_0(\mathbf q,\omega_n)} ,
\label{eq:SI_stationarity_static}
\end{equation}
which is \cref{eq:S_reconstructed}. For a measured $S(\mathbf q)$ and a known $S_0(\mathbf q,\omega_n)$, this is one equation for the single unknown $\lambda_{\mathbf q}$ of each $\mathbf q$. Its right-hand side equals $S_0(\mathbf q)$ at $\lambda_{\mathbf q}=0$ [\cref{eq:SI_static_from_Matsubara}] and decreases monotonically with $\lambda_{\mathbf q}$, since every term does. As $\lambda_{\mathbf q}\to\infty$, it tends to zero. As $\lambda_{\mathbf q}$ decreases to $-1/[2NS_0(\mathbf q,0)]$, it diverges, first in the term $n=0$, where $S_0(\mathbf q,\omega_n)$ is largest (\cref{SI:spectral}). \Cref{eq:SI_stationarity_static} therefore has a single root in this range for any $S(\mathbf q)>0$, positive if $S(\mathbf q)<S_0(\mathbf q)$ and negative otherwise. A bracketing root search finds it. In practice the sum in \cref{eq:SI_stationarity_static} converges slowly: its terms fall off as $\omega_n^{-2}$, like $S_0(\mathbf q,\omega_n)$ [\cref{eq:SI_tail}], so cutting it at $|n|\leq n_c$ leaves an error of order $1/n_c$. We can rewrite the sum to improve convergence. We add and subtract $S_0(\mathbf q,\omega_n)$ in each term,
\begin{equation}
S_0(\mathbf q,\omega_n)+\frac{S_0(\mathbf q,\omega_n)}{1+2N\lambda_{\mathbf q}S_0(\mathbf q,\omega_n)}-S_0(\mathbf q,\omega_n)=S_0(\mathbf q,\omega_n)-\frac{2N\lambda_{\mathbf q}S_0^2(\mathbf q,\omega_n)}{1+2N\lambda_{\mathbf q}S_0(\mathbf q,\omega_n)} ,
\end{equation}
where the last term falls off as $\omega_n^{-4}$. With $\beta^{-1}\sum_nS_0(\mathbf q,\omega_n)=S_0(\mathbf q)$, the static structure factor of the reference [\cref{eq:SI_static_from_Matsubara}], \cref{eq:SI_stationarity_static} therefore becomes
\begin{equation}
S(\mathbf q)=S_0(\mathbf q)-\frac{1}{\beta}\sum_n\frac{2N\lambda_{\mathbf q}S_0^2(\mathbf q,\omega_n)}{1+2N\lambda_{\mathbf q}S_0(\mathbf q,\omega_n)} ,
\end{equation}
and cutting this sum at $|n|\leq n_c$ leaves an error of order $n_c^{-3}$.

With this root,
\begin{equation}
S_\lambda(\mathbf q,\omega_n)\equiv\frac{S_0(\mathbf q,\omega_n)}{1+2N\lambda_{\mathbf q}S_0(\mathbf q,\omega_n)}
\label{eq:SI_S_lambda}
\end{equation}
is the correlation function of the inferred state in the Gaussian approximation, and \cref{eq:SI_stationarity_static} requires it to reproduce $S(\mathbf q)$ only. Substituting $1+2N\lambda_{\mathbf q}S_0=S_0/S_\lambda$ and, by \cref{eq:SI_stationarity_static}, $N\beta\lambda_{\mathbf q}S(\mathbf q)=\sum_nN\lambda_{\mathbf q}S_\lambda=\frac12\sum_n(1-S_\lambda/S_0)$ into \cref{eq:SI_Hrel_lambda_static} gives
\begin{equation}
H_{\text{rel}}\approx\frac12\sum_{\mathbf q,n}\left[\ln\frac{S_\lambda(\mathbf q,\omega_n)}{S_0(\mathbf q,\omega_n)}-\frac{S_\lambda(\mathbf q,\omega_n)}{S_0(\mathbf q,\omega_n)}+1\right] ,
\label{eq:SI_Hrel_static}
\end{equation}
which is \cref{eq:H_rel_main} with $S(\mathbf q,\omega_n)$ replaced by $S_\lambda(\mathbf q,\omega_n)$.

\subsection{Exact inference}
\label{SI:exact}

The exact inference is a numerical search for the stationary point of $H_{\text{rel}}(\lambda)\equiv\ln\mathcal Z(\lambda)-\ln\mathcal I[0]$ over the multipliers $\lambda_{\mathbf q,n}$. With $P_0[R]=e^{-\mathcal A_0[R]}/\mathcal I[0]$, \cref{eq:effectiveZ_PI} writes this function as an average over the paths of the ideal gas,
\begin{equation}
H_{\text{rel}}(\lambda)=N\sum_{\mathbf q,n}\lambda_{\mathbf q,n}S(\mathbf q,\omega_n)+\ln\Big\langle\exp\Big(-\sum_{\mathbf q,n}\lambda_{\mathbf q,n}|\rho_{\mathbf q,n}[R]|^2\Big)\Big\rangle_0 .
\label{eq:SI_Hrel_numerical}
\end{equation}
By \cref{eq:SI_dlnZ}, the stationary point is where the inferred weight reproduces the given correlations, and by \cref{eq:SI_Hrel_lnZ} the value there is $H_{\text{rel}}$. For bosons and distinguishable particles, whose path weights are probabilities, the derivatives of \cref{eq:SI_dlnZ} with respect to the multipliers are the covariances of the $|\rho_{\mathbf q,n}[R]|^2$ in $P[R]$, a positive semidefinite matrix, so $H_{\text{rel}}(\lambda)$ is convex and the search is a minimization. For fermions the path weights change sign, so this matrix can have negative eigenvalues, and the inference is defined by stationarity alone (\cref{SI:maximization}). At the stationary point, $F_{\text{ex}}$, the kinetic shift and $H_{\text{ex}}$ follow from the exact relations of \cref{SI:legendre}.

There is a way to simplify the numerical optimization by using additional information: the interaction is an instantaneous pair interaction $U_{\mathbf q}$. The exact multipliers then do not depend on $n$, and $P[R]$ is the path representation of $\hat P_\lambda$ (\cref{SI:n_independent}), so the path average in \cref{eq:SI_Hrel_numerical} is $\mathcal Z_\lambda/\mathcal I[0]$. With $\sum_nS(\mathbf q,\omega_n)=\beta S(\mathbf q)$ [\cref{eq:SI_static_from_Matsubara}], \cref{eq:SI_Hrel_numerical} becomes
\begin{equation}
H_{\text{rel}}(\lambda)=N\beta\sum_{\mathbf q}\lambda_{\mathbf q}S(\mathbf q)+\ln\mathcal Z_\lambda-\ln\mathcal I[0] .
\label{eq:SI_Hrel_instantaneous}
\end{equation}
The optimization thus runs over one multiplier per wave vector and needs only $S(\mathbf q)$. In this case, the kinetic term of the reference, $\hat{\mathcal H}_0$, supplies the imaginary-time information missing from $S(\mathbf q)$.

\section{Correlation free energy and the reversed relative entropy}
\label{SI:reversed}

For two thermal states, $\hat{P}=e^{-\beta(\hat{\mathcal H}-F)}$ and $\hat{P}_0=e^{-\beta(\hat{\mathcal H}_0-F_0)}$, with $\langle\cdot\rangle_0=\mathrm{tr}(\hat P_0\,\cdot)$,
\begin{equation}
    D(\hat{P}_0\|\hat{P}) = \text{tr}\left(\hat{P}_0\ln\hat{P}_0-\hat{P}_0\ln\hat{P}\right)
    = \beta\langle\hat{\mathcal H}-\hat{\mathcal H}_0\rangle_0 - \beta(F-F_0) .
\label{eq:SI_D_general}
\end{equation}
The correlation free energy, the part of $F-F_0$ beyond first order in $\hat{\mathcal H}-\hat{\mathcal H}_0$, is therefore
\begin{equation}
    \Delta F_{\text{corr}} \equiv (F-F_0) - \langle\hat{\mathcal H}-\hat{\mathcal H}_0\rangle_0 = -\beta^{-1}D(\hat{P}_0\|\hat{P}) .
\label{eq:SI_Fcorr_general}
\end{equation}
Since $D(\hat P_0\|\hat P)\ge0$, $\Delta F_{\text{corr}}\le0$; this is the Gibbs--Bogoliubov inequality $F\le F_0+\langle\hat{\mathcal H}-\hat{\mathcal H}_0\rangle_0$.

For the inferred state, $F-F_0=F_{\text{ex}}$, and by \cref{eq:Zpath,eq:Fex_def},
\begin{equation}
    \beta F_{\text{ex}}=-\ln\frac{\mathcal Z_\lambda}{\mathcal I[0]}
    =-\ln\frac{\oint\mathcal D\mathbf R\,\exp\big(-\mathcal A_0[R]-\sum_{\mathbf q,n}\lambda_{\mathbf q,n}|\rho_{\mathbf q,n}[R]|^2\big)}{\oint\mathcal D\mathbf R\,\exp\big(-\mathcal A_0[R]\big)}
    =-\ln\Big\langle\exp\Big(-\sum_{\mathbf q,n}\lambda_{\mathbf q,n}|\rho_{\mathbf q,n}[R]|^2\Big)\Big\rangle_0 .
\label{eq:SI_Fex_paths}
\end{equation}
Its first-order term in the multipliers is $\sum_{\mathbf q,n}\lambda_{\mathbf q,n}\langle|\rho_{\mathbf q,n}[R]|^2\rangle_0=N\sum_{\mathbf q,n}\lambda_{\mathbf q,n}S_0$ [\cref{eq:SI_Matsubara_mode_covariance}].

By \cref{eq:Fex,eq:lambda},
\begin{equation}
    \beta F_{\text{ex}}\approx\frac12\sum_{\mathbf q,n}\ln\frac{S_0}{S}=\frac12\sum_{\mathbf q,n}\ln\big(1+2N\lambda_{\mathbf q,n}S_0\big) .
\label{eq:SI_Fex_first_order}
\end{equation}
Subtracting the first-order term [\cref{eq:SI_Fcorr_general}] gives
\begin{equation}
    \beta\Delta F_{\text{corr}} = \beta F_{\text{ex}} - N\sum_{\mathbf q,n}\lambda_{\mathbf q,n}S_0 \approx \frac{1}{2}\sum_{\mathbf q,n}\left[\ln\frac{S_0}{S}-\frac{S_0}{S}+1\right] ,
\label{eq:SI_Fcorr_explicit}
\end{equation}
which is \cref{eq:Fcorr}: the summand of \cref{eq:H_rel_main} with $S/S_0$ replaced by $S_0/S$, and therefore non-positive term by term.

\section{Reference correlation functions}
\label{sec:reference}

We derive the density-density correlation function of the reference, $S_0(\mathbf{q},\omega_n)$, for the ideal Bose and Fermi gases used in the applications. The reference correlation function is \cref{eq:SI_Matsubara_transform} with $\hat{P}\to\hat{P}_0$, and with $\hat{\mathcal{H}}\to\hat{\mathcal{H}}_0$ in the imaginary-time evolution.

\subsection{Ideal Bose gas below the condensation temperature}
\label{SI:ref_bose}

In second quantization, with $\hat{\psi}(\mathbf{r})=V^{-1/2}\sum_{\mathbf{k}}e^{i\mathbf{k}\cdot\mathbf{r}}\hat{a}_{\mathbf{k}}$ and $\int\mathrm{d}\mathbf{r}\,e^{i(\mathbf{k}'-\mathbf{k}-\mathbf{q})\cdot\mathbf{r}}=V\delta_{\mathbf{k}',\mathbf{k}+\mathbf{q}}$, the density mode of \cref{eq:SI_density_mode} is
\begin{equation}
    \hat{\rho}_{\mathbf{q}}=\int\mathrm{d}\mathbf{r}\,e^{-i\mathbf{q}\cdot\mathbf{r}}\,\hat{\psi}^{\dagger}(\mathbf{r})\hat{\psi}(\mathbf{r})=\sum_{\mathbf{k}}\hat{a}^{\dagger}_{\mathbf{k}}\hat{a}_{\mathbf{k}+\mathbf{q}} .
\end{equation}
Below the condensation temperature the mode $\mathbf{k}=\mathbf{0}$ holds $N_0$ particles. Two terms of the sum involve the condensate, $\mathbf{k}=\mathbf{0}$ and $\mathbf{k}=-\mathbf{q}$. Keeping only these and replacing $\hat{a}_{\mathbf{0}}$ and $\hat{a}^{\dagger}_{\mathbf{0}}$ by $\sqrt{N_0}$,
\begin{equation}
    \hat{\rho}_{\mathbf{q}}\simeq\sqrt{N_0}\,\big(\hat{a}_{\mathbf{q}}+\hat{a}^{\dagger}_{-\mathbf{q}}\big) .
\label{eq:SI_rho_condensate}
\end{equation}
This neglects the terms bilinear in the non-condensed modes, and is accurate when the depletion $1-N_0/N$ is small. We use the grand-canonical reference, whose chemical potential, below the condensation temperature and in the thermodynamic limit, tends to zero, so $\hat{\mathcal{H}}_0=\sum_{\mathbf{k}}\epsilon_k\hat{a}^{\dagger}_{\mathbf{k}}\hat{a}_{\mathbf{k}}$. The operators $\hat a_{\mathbf q}$ and $\hat a^\dagger_{-\mathbf q}$ change the energy by $\mp\epsilon_q$, so the matrix elements in \cref{eq:SI_spectral} vanish unless $E_l-E_m=\pm\epsilon_q$, and $S_0(\mathbf q,\omega_n)$ is a single pole at $\epsilon_q$. By \cref{eq:SI_tail}, its numerator is $M_1$, which is $2\epsilon_q$ by the f-sum rule, \cref{eq:SI_fsum}, up to corrections of the order of the depletion. Hence
\begin{equation}
    S_0(\mathbf{q},\omega_n)=\frac{2\epsilon_q}{(\hbar\omega_n)^2+\epsilon_q^2} .
\label{eq:S0_bose}
\end{equation}

\subsection{Ideal Fermi gas}
\label{SI:ref_fermi}

For the ideal Fermi gas, $S_0(\mathbf q,\omega_n)$ is $-\bar\rho^{-1}$ times the finite-temperature Lindhard function. For non-interacting fermions with $g_s$ spin states ($g_s=2$ in the electron gas), $\hat{\rho}_{\mathbf{q}}=\sum_{\mathbf{k}\sigma}\hat{c}^{\dagger}_{\mathbf{k}\sigma}\hat{c}_{\mathbf{k}+\mathbf{q},\sigma}$. In the grand-canonical reference at chemical potential $\mu$, $\hat{c}^{(\tau)}_{\mathbf{k}\sigma}=e^{-\tau(\epsilon_k-\mu)/\hbar}\hat{c}_{\mathbf{k}\sigma}$ and $\hat{c}^{\dagger(\tau)}_{\mathbf{k}\sigma}=e^{\tau(\epsilon_k-\mu)/\hbar}\hat{c}^{\dagger}_{\mathbf{k}\sigma}$, so $\hat{\rho}_{\mathbf{q}}^{(\tau)}=\sum_{\mathbf{k}\sigma}e^{-\tau(\epsilon_{k+q}-\epsilon_k)/\hbar}\hat{c}^{\dagger}_{\mathbf{k}\sigma}\hat{c}_{\mathbf{k}+\mathbf{q},\sigma}$, where $\epsilon_{k+q}\equiv\epsilon_{|\mathbf{k}+\mathbf{q}|}$. For $\mathbf{q}\neq0$, Wick's theorem leaves the single contraction that returns each fermion to its initial momentum,
\begin{equation}
    \big\langle\hat{\rho}_{\mathbf{q}}^{(\tau)}\hat{\rho}_{-\mathbf{q}}^{(0)}\big\rangle_0=\sum_{\mathbf{k}\sigma}e^{-\tau(\epsilon_{k+q}-\epsilon_k)/\hbar}\,f_k\,(1-f_{k+q}),\qquad f_k=\frac{1}{e^{\beta(\epsilon_k-\mu)}+1} .
\end{equation}
We apply \cref{eq:SI_tau_integral} with $a=\epsilon_{k+q}-\epsilon_k$ and use the detailed-balance identity $f_k(1-f_{k+q})\big(1-e^{-\beta(\epsilon_{k+q}-\epsilon_k)}\big)=f_k-f_{k+q}$:
\begin{equation}
    S_0(\mathbf{q},\omega_n)=\frac{1}{N}\sum_{\mathbf{k}\sigma}\frac{f_k-f_{k+q}}{\epsilon_{k+q}-\epsilon_k-i\hbar\omega_n}
    =-\frac{1}{\bar{\rho}}\,\chi_0(\mathbf{q},i\omega_n),
\label{eq:S0_fermi}
\end{equation}
with $\mu$ fixed by $\sum_{\mathbf{k}\sigma}f_k=N$. Terms with $\epsilon_{k+q}=\epsilon_k$ stand for their limit, $\beta f_k(1-f_k)\delta_{n0}$.

\section{Single-pole correlation functions and the Bogoliubov gas}
\label{sec:harmonic}

In \cref{SI:harmonic} we evaluate the closed forms for single-pole correlation functions, which describe harmonic oscillators. In \cref{sec:weakbose} we apply them to the weakly interacting Bose gas within the Bogoliubov approximation and compare the result with its exact excess entropy. In \cref{SI:equal_time} we evaluate the classical relation with the static structure factor of the same gas, compare it with the exact excess entropy, and apply the inference of \cref{SI:n_independent} to the same static structure factor.

\subsection{Single-pole correlation functions}
\label{SI:harmonic}

Let the reference and the interacting correlation functions be single poles, mode by mode,
\begin{equation}
    S_0(\mathbf{q},\omega_n)=\frac{2\epsilon_q}{(\hbar\omega_n)^2+a_q^2},\qquad
    S(\mathbf{q},\omega_n)=\frac{2\epsilon_q}{(\hbar\omega_n)^2+b_q^2},
\label{eq:SI_single_pole}
\end{equation}
with $a_q,b_q>0$ independent of temperature. The numerator is $2\epsilon_q$ for both by the f-sum rule, \cref{eq:SI_fsum}. \Cref{eq:lambda} then gives
\begin{equation}
    \lambda_{\mathbf{q},n}=\frac{(\hbar\omega_n)^2+b_q^2-(\hbar\omega_n)^2-a_q^2}{4N\epsilon_q}=\frac{b_q^2-a_q^2}{4N\epsilon_q} .
\label{eq:SI_lambda_single_pole}
\end{equation}

By \cref{eq:Fex}, mode $\mathbf{q}$ contributes to $\beta F_{\text{ex}}$
\begin{multline}
    \beta F_{\mathbf{q}}=\frac{1}{2}\sum_{n}\ln\frac{(\hbar\omega_n)^2+b_q^2}{(\hbar\omega_n)^2+a_q^2}\\
    =\ln\frac{b_q}{a_q}+\sum_{n=1}^{\infty}\ln\Big(1+\frac{b_q^2}{(\hbar\omega_n)^2}\Big)-\sum_{n=1}^{\infty}\ln\Big(1+\frac{a_q^2}{(\hbar\omega_n)^2}\Big) ,
\end{multline}
where the term $n=0$ gives $\ln(b_q/a_q)$ and the terms $\pm n$ are equal. With $\hbar\omega_n=2\pi n/\beta$, the product formula $\sinh y=y\prod_{n=1}^{\infty}[1+y^2/(n\pi)^2]$ at $y=\beta b_q/2$ and $y=\beta a_q/2$ gives
\begin{equation}
    \beta F_{\mathbf{q}}
    =\ln\frac{b_q}{a_q}+\ln\frac{\sinh(\beta b_q/2)}{\beta b_q/2}-\ln\frac{\sinh(\beta a_q/2)}{\beta a_q/2}
    =\ln\frac{\sinh(\beta b_q/2)}{\sinh(\beta a_q/2)}
    =\frac{\beta(b_q-a_q)}{2}+\ln\frac{1-e^{-\beta b_q}}{1-e^{-\beta a_q}} ,
\label{eq:SI_F_single_pole}
\end{equation}
where the last form, from $\sinh y=\tfrac{1}{2}e^{y}(1-e^{-2y})$, is the free energy of a harmonic oscillator of energy $b_q$, zero-point energy included, minus that of an oscillator of energy $a_q$. At fixed multipliers, $b_q^2=a_q^2+4N\epsilon_q\lambda_{\mathbf q,n}$ [\cref{eq:SI_lambda_single_pole}] does not depend on temperature, so the Legendre transform, \cref{eq:SI_Legendre}, gives the difference of their entropies,
\begin{equation}
    H_{\mathbf{q}}=s(\beta b_q)-s(\beta a_q),\qquad s(y)\equiv\frac{y}{e^{y}-1}-\ln\big(1-e^{-y}\big),
\label{eq:SI_H_single_pole}
\end{equation}
with $s$ the entropy of a harmonic oscillator.

\subsection{Weakly interacting Bose gas}
\label{sec:weakbose}
\label{SI:bogoliubov}

The Hamiltonian is
\begin{equation}
    \hat{\mathcal{H}}=\sum_{\mathbf{q}}\epsilon_q\,\hat{a}_{\mathbf q}^{\dagger}\hat{a}_{\mathbf q}
    +\frac{g}{2V}\sum_{\mathbf{k}\mathbf{k}'\mathbf{q}}
    \hat{a}_{\mathbf q-\mathbf k'}^{\dagger}\hat{a}_{\mathbf k+\mathbf k'}^{\dagger}
    \hat{a}_{\mathbf k}\hat{a}_{\mathbf q},
\end{equation}
with contact coupling $g>0$, so that $U_{\mathbf q}=g$. In the Bogoliubov approximation~\cite{pitaevskii2016bose}, $\hat{a}_{\mathbf{0}}$ is replaced by $\sqrt{N_0}$, terms beyond second order in the non-condensed modes are dropped, and $N_0$ is replaced by $N$ to leading order in the depletion. The transformation $\hat{a}_{\mathbf{q}}=u_q\hat{\alpha}_{\mathbf{q}}-v_q\hat{\alpha}^{\dagger}_{-\mathbf{q}}$, with $u_q^2-v_q^2=1$, preserves the bosonic commutators and, for suitable $u_q$ and $v_q$, diagonalizes the Hamiltonian, $\hat{\mathcal{H}}\simeq E_{\text{gs}}+\sum_{\mathbf{q}\neq0}E_q\hat{\alpha}_{\mathbf{q}}^{\dagger}\hat{\alpha}_{\mathbf{q}}$, with quasiparticle energies
\begin{equation}
    E_q=\sqrt{\epsilon_q^2+2\bar{\rho}g\,\epsilon_q}
\label{eq:bogoliubov_spectrum}
\end{equation}
and a ground-state energy $E_{\text{gs}}$ that does not depend on temperature. By \cref{eq:SI_rho_condensate}, $\hat\rho_{\mathbf q}$ is then linear in $\hat\alpha_{\mathbf q}$ and $\hat\alpha^\dagger_{-\mathbf q}$, which change the energy by $\mp E_q$. As for the reference (\cref{SI:ref_bose}), $S(\mathbf q,\omega_n)$ is then a single pole at $E_q$ with numerator $2\epsilon_q$, which is \cref{eq:SI_single_pole} with $a_q=\epsilon_q$ [\cref{eq:S0_bose}] and $b_q=E_q$. With $E_q^2-\epsilon_q^2=2\bar{\rho}g\,\epsilon_q$ and $\bar{\rho}=N/V$, \cref{eq:SI_lambda_single_pole} gives
\begin{equation}
    \lambda_{\mathbf q,n}=\frac{E_q^2-\epsilon_q^2}{4N\epsilon_q}=\frac{g}{2V},
\label{eq:lambda_weakbose}
\end{equation}
so $2V\lambda_{\mathbf q,n}=U_{\mathbf q}=g$, the contact interaction of the Hamiltonian.

By \cref{eq:SI_F_single_pole,eq:SI_H_single_pole}, the closed forms give
\begin{equation}
    \beta F_{\text{ex}} = \sum_{\mathbf{q}\neq0}\ln\frac{\sinh(\beta E_q/2)}{\sinh(\beta\epsilon_q/2)}
\label{eq:Fex_bose}
\end{equation}
and
\begin{equation}
    H_{\text{ex}}^{\text{Bog}}=\sum_{\mathbf{q}\neq0}\big[s(\beta E_q)-s(\beta\epsilon_q)\big] .
\label{eq:Hex_exact_bose}
\end{equation}
The entropies of the Bogoliubov gas and of the condensed ideal gas are those of their independent modes, $\sum_{\mathbf{q}\neq0}s(\beta E_q)$ and $\sum_{\mathbf{q}\neq0}s(\beta\epsilon_q)$, since the condensate carries no entropy, so \cref{eq:Hex_exact_bose} is the exact excess entropy. Their free energies are $F_{\text{Bog}}=E_{\text{gs}}+\beta^{-1}\sum_{\mathbf{q}\neq0}\ln(1-e^{-\beta E_q})$ and $F_0=\beta^{-1}\sum_{\mathbf{q}\neq0}\ln(1-e^{-\beta\epsilon_q})$, so by the last form of \cref{eq:SI_F_single_pole}
\begin{equation}
    F_{\text{ex}}-\big(F_{\text{Bog}}-F_0\big)=\frac{1}{2}\sum_{\mathbf{q}\neq0}(E_q-\epsilon_q)-E_{\text{gs}} ,
\end{equation}
a temperature-independent constant. The agreement is exact to the order at which the single-pole form of $S(\mathbf q,\omega_n)$ is exact, the leading order in the depletion.

The density modes of the reference are linear in the coordinates of harmonic oscillators [\cref{eq:SI_rho_condensate}], whose path weight is Gaussian. The density modes are therefore Gaussian, and the Gaussian approximation is exact (\cref{SI:closed_form}).

\subsection{The static structure factor}
\label{SI:equal_time}

By \cref{eq:SI_static_from_Matsubara}, with $\hbar\omega_n=2\pi n/\beta$ and $\sum_n(n^2+x^2)^{-1}=(\pi/x)\coth(\pi x)$, the static structure factor of the gas is
\begin{equation}
    S(q)=\frac{1}{\beta}\sum_n\frac{2\epsilon_q}{(\hbar\omega_n)^2+E_q^2}=\frac{\epsilon_q}{E_q}\coth\frac{\beta E_q}{2} ,
\end{equation}
and that of the reference, with $E_q$ replaced by $\epsilon_q$, is $S_0(q)=\coth(\beta\epsilon_q/2)$. The classical relation, \cref{eq:classic_entropy_estimation}, evaluated with the equal-time ratio $r_q=S(q)/S_0(q)$, gives
\begin{equation}
    H_{\text{static}}=\frac{1}{2}\sum_{\mathbf q\neq0}\big(\ln r_q-r_q+1\big) .
\end{equation}
In the thermodynamic limit, $\sum_{\mathbf q}\to(V/2\pi^2)\int\mathrm dq\,q^2$. As $\beta\bar\rho g$ grows from $0.05$ to $20$, $|H_{\text{ex}}^{\text{Bog}}|$ [\cref{eq:Hex_exact_bose}] falls by a factor of about $400$ and $|H_{\text{static}}|$ by a factor of $3$, so $H_{\text{static}}/H_{\text{ex}}^{\text{Bog}}$ grows from $0.1$ to $13$ (\cref{fig:bose}).

At zero temperature all modes are in their ground state, which carries no entropy, so $H_{\text{ex}}^{\text{Bog}}\to0$. Both $\coth$ factors tend to unity, so $r_q\to\epsilon_q/E_q$, which is below unity by \cref{eq:bogoliubov_spectrum}, and every term of $H_{\text{static}}$ stays negative. At $\beta\bar\rho g=20$, $H_{\text{static}}$ is within $3\%$ of this zero-temperature limit.

The multipliers of this gas, \cref{eq:lambda_weakbose}, do not depend on $n$, so the inference of \cref{SI:n_independent} applies. With them, $S_\lambda$ of \cref{eq:SI_S_lambda} is $S(\mathbf q,\omega_n)$ itself [\cref{eq:lambda}], whose sum over $n$ is $\beta S(q)$, so they solve \cref{eq:SI_stationarity_static}. Since \cref{eq:SI_stationarity_static} has a single root, the inference of \cref{SI:n_independent} returns these multipliers, the full $S(\mathbf q,\omega_n)$ and $H_{\text{ex}}^{\text{Bog}}$ [\cref{eq:Hex_exact_bose}] from $S(q)$ and $S_0(\mathbf q,\omega_n)$ alone.

\section{Details of the spinless-fermion benchmark}
\label{SI:benchmark}

\subsection{Model, density modes and exact correlation functions}
\label{SI:lehmann}

We use spinless fermions. With spin, part of the entropy would be in the spin configurations, which the density correlations do not resolve. The density modes $\hat\rho_q$ are those of \cref{sec:lattice}, and numerical values use $\hbar=t=1$. We diagonalize \cref{eq:tW_model} in the sector of fixed particle number $N$, of dimension $\binom{L}{N}$ ($924$ for $L=12$, $N=6$). Its eigenstates, $\hat{\mathcal{H}}|m\rangle=E_m|m\rangle$, give $\hat P=\sum_m p_m|m\rangle\langle m|$, with $p_m=e^{-\beta E_m}/Z$ and $Z=\sum_m e^{-\beta E_m}$. The reference is the free ring, $W=0$, in the same sector, so $\hat\rho_0=N$ in both $\hat P$ and $\hat P_0$.

\Cref{eq:tW_model} is real in the occupation-number basis, so it is invariant under time reversal, as \cref{SI:density_correlations} assumes. The spectral representation, \cref{eq:SI_spectral}, with these eigenstates gives the exact $S(q,\omega_n)$, and with those at $W=0$ it gives $S_0(q,\omega_n)$.

\subsection{Exact pair-density control}
\label{SI:control}

The numerical inference shown by the dots in \cref{fig:benchmark} is a control for the closed form: it uses no Gaussian approximation and returns $\hat P$ itself. It takes the multipliers independent of $n$ and searches for the stationary point of \cref{eq:SI_Hrel_instantaneous} over the $\lambda_q$, with $\hat P_\lambda$ of \cref{SI:n_independent} evaluated by diagonalization. Since $\hat\rho_q\hat\rho_{-q}=\hat\rho_{-q}\hat\rho_q$, only $\lambda_q+\lambda_{-q}$ enters, and we take $\lambda_{-q}=\lambda_q$. This leaves the $L/2$ parameters $\lambda_q$ with $k=1,\dots,L/2$. Summed over all $L$ wave numbers, including $q=0$, $\sum_q e^{-iq(j-l)}=L\delta_{jl}$, so $\sum_q\hat\rho_q\hat\rho_{-q}=L\sum_j\hat n_j^2$. For spinless fermions $\hat n_j^2=\hat n_j$, so this is $LN$, and the sum over $q\neq0$ is $LN-N^2$, a constant in the sector of fixed $N$. A common shift of the $\lambda_q$ therefore adds a constant to the exponent of $\hat P_\lambda$, which cancels against $\mathcal Z_\lambda$, so the multipliers are fixed only up to this shift.

The search is the minimization of a convex function, and its minimum is the constrained maximum of $H_{\mathrm{rel}}$. For a density matrix $\hat P$ with $\mathrm{tr}(\hat P\hat\rho_q\hat\rho_{-q})=NS(q)$ for every $q\neq0$, and for arbitrary $\lambda_q$, \cref{eq:H_rel_def,eq:SI_Hrel_instantaneous} give
\begin{equation}
    H_{\mathrm{rel}}[\hat P]+D(\hat P\|\hat P_\lambda)
    =\mathrm{tr}\big[\hat P\big(\ln\hat P_0-\ln\hat P_\lambda\big)\big]
    =H_{\mathrm{rel}}(\lambda).
\label{eq:SI_dual_identity}
\end{equation}
By Klein's inequality~\cite{vedral2002role}, $D(\hat P\|\hat P_\lambda)\geq0$, with equality only for $\hat P=\hat P_\lambda$, so $H_{\mathrm{rel}}[\hat P]\leq H_{\mathrm{rel}}(\lambda)$ for every such $\hat P$ and every $\lambda$. If $\hat P_\lambda$ has the given $S(q)$ at some $\lambda$, the bound is attained there, $H_{\mathrm{rel}}=\min_\lambda H_{\mathrm{rel}}(\lambda)$, and $\hat P_\lambda$ is the unique maximizer. For every density matrix $\hat P$, the same inequality gives $\ln\mathcal Z_\lambda\geq-\mathrm{tr}\,\hat P\ln\hat P-\beta\,\mathrm{tr}(\hat P\hat{\mathcal H}_0)-\beta\sum_{q\neq0}\lambda_q\,\mathrm{tr}(\hat P\hat\rho_q\hat\rho_{-q})$, with equality at $\hat P=\hat P_\lambda$. Thus $\ln\mathcal Z_\lambda$ is a maximum of functions affine in $\lambda$, $H_{\mathrm{rel}}(\lambda)$ is convex, and any local minimum is global.

By \cref{eq:U_lattice}, $\hat P_\lambda=\hat P$ at $\lambda_q=U_q/2L=W\cos q/L$, since the constant $WN^2/L$ cancels in the normalization. $\hat P$ has the given $S(q)$, so it is the maximizer. The search therefore returns $\hat P$ and the exact entropy, with $\lambda_q=W\cos q/L$ up to the common shift. The density matrices that reproduce every $S(q,\omega_n)$ also reproduce $S(q)$ [\cref{eq:SI_static_from_Matsubara}], and $\hat P$ is among them, so $\hat P$ also solves the inference with the full set of constraints. The minimization reproduces $S(q)$ to within $10^{-8}$ and $H=-\sum_m p_m\ln p_m$ to a relative deviation below $5\times10^{-9}$, set by the optimizer tolerance.

\subsection{Lattice high-frequency sum rule}
\label{SI:sumrule}

On the lattice, the kinetic energy sets the high-frequency tail of $S(q,\omega_n)$. By \cref{eq:SI_tail,eq:SI_highfreq}, $\omega_n^2S(q,\omega_n)\to M_1/\hbar^2$ at large $|n|$, with $M_1=N^{-1}\langle[\hat{\rho}_q,[\hat{\mathcal{H}},\hat{\rho}_{-q}]]\rangle$. The interaction commutes with $\hat\rho_{-q}$, since both are functions of the $\hat n_j$, so only the hopping term, $\hat{\mathcal H}_0=-t\sum_k(\hat c_k^\dagger\hat c_{k+1}+\hat c_{k+1}^\dagger\hat c_k)$, contributes. With $[\hat n_j,\hat c_k^\dagger]=\delta_{jk}\hat c_k^\dagger$ and $[\hat n_j,\hat c_k]=-\delta_{jk}\hat c_k$,
\begin{equation}
    [\hat{n}_j,\hat{c}^{\dagger}_k\hat{c}_{k+1}]=(\delta_{jk}-\delta_{j,k+1})\,\hat{c}^{\dagger}_k\hat{c}_{k+1},
    \qquad
    [\hat{n}_j,\hat{c}^{\dagger}_{k+1}\hat{c}_{k}]=-(\delta_{jk}-\delta_{j,k+1})\,\hat{c}^{\dagger}_{k+1}\hat{c}_{k}.
\end{equation}
Multiplying by $e^{iqj}$ and summing over $j$, with $[\hat{\mathcal H}_0,\hat\rho_{-q}]=-\sum_je^{iqj}[\hat n_j,\hat{\mathcal H}_0]$,
\begin{equation}
    \big[\hat{\mathcal{H}}_0,\hat{\rho}_{-q}\big]
    = t\sum_k e^{iqk}\big(1-e^{iq}\big)
      \big(\hat{c}_k^{\dagger}\hat{c}_{k+1}-\hat{c}_{k+1}^{\dagger}\hat{c}_k\big).
\end{equation}
The same two commutators, multiplied by $e^{-iqj}$, give $[\hat\rho_q,\hat c_k^\dagger\hat c_{k+1}]=e^{-iqk}(1-e^{-iq})\,\hat c_k^\dagger\hat c_{k+1}$ and $[\hat\rho_q,\hat c_{k+1}^\dagger\hat c_{k}]=-e^{-iqk}(1-e^{-iq})\,\hat c_{k+1}^\dagger\hat c_{k}$. With $(1-e^{iq})(1-e^{-iq})=2(1-\cos q)$,
\begin{equation}
    \big[\hat{\rho}_q,[\hat{\mathcal{H}},\hat{\rho}_{-q}]\big]
    = 2t(1-\cos q)\sum_k\big(\hat{c}_k^{\dagger}\hat{c}_{k+1}+\hat{c}_{k+1}^{\dagger}\hat{c}_k\big)
    = -2(1-\cos q)\,\hat{\mathcal{H}}_0 .
\label{eq:SI_commutator}
\end{equation}
Its thermal average gives $M_1=-2(1-\cos q)E_{\mathrm{kin}}/N$. Since $M_1\geq0$ [\cref{eq:SI_tail}], $E_{\mathrm{kin}}\leq0$, and $\omega_n^2S\to2(1-\cos q)|E_{\mathrm{kin}}|/N\hbar^2$, which is \cref{eq:tail_lattice}.

\subsection{Rescaled reference}
\label{SI:bridge}

The rescaled reference makes the Matsubara sum of \cref{eq:H_rel_main} converge on the lattice. Its correlation function $S_0^{\gamma}(q,\omega_n)$ is \cref{eq:SI_spectral} with the eigenstates of $\gamma\hat{\mathcal H}_0$ [\cref{eq:P0gamma}], that is, with $t\to\gamma t$. The summand of \cref{eq:H_rel_main}, $h(x)=\tfrac12(\ln x-x+1)$ with $x=S/S_0^{\gamma}$, vanishes only at $x=1$, so the Matsubara sum converges only if $S/S_0^{\gamma}\to1$ as $|n|\to\infty$. \Cref{eq:tail_lattice} for the free ring with hopping $\gamma t$ gives $\omega_n^2S_0^{\gamma}\to2(1-\cos q)|E^0_{\mathrm{kin}}(\gamma)|/N\hbar^2$, so
\begin{equation}
    \frac{S(q,\omega_n)}{S_0^{\gamma}(q,\omega_n)}\;\xrightarrow{|n|\to\infty}\;\frac{E_{\mathrm{kin}}}{E^0_{\mathrm{kin}}(\gamma)} .
\label{eq:SI_ratio_limit}
\end{equation}
The factor $2(1-\cos q)$ has cancelled, so the single condition $E_{\mathrm{kin}}^{0}(\gamma^{\ast})=E_{\mathrm{kin}}$, \cref{eq:gammastar}, makes the limit unity for every $q$. The two correlation functions then differ first at order $\omega_n^{-4}$, so $x-1=O(\omega_n^{-2})$. Since $h(x)=-\tfrac14(x-1)^2+O\big((x-1)^3\big)$, the summand decays as $\omega_n^{-4}$, and the sum converges.

An exact relation then converts the result to the entropy. By \cref{eq:H_rel_def}, with $\ln\hat{P}_0^{\gamma}=-\beta\gamma\hat{\mathcal{H}}_0-\ln\mathcal{I}[0;\gamma]$ [\cref{eq:P0gamma}],
\begin{equation}
    H_{\mathrm{rel}}(\gamma) = H - \beta\gamma E_{\mathrm{kin}} - \ln\mathcal{I}[0;\gamma],
\label{eq:SI_Hrel_gamma}
\end{equation}
where $E_{\mathrm{kin}}=\operatorname{tr}(\hat{P}\hat{\mathcal{H}}_0)$ does not depend on $\gamma$. We solve \cref{eq:SI_Hrel_gamma} at $\gamma^{\ast}$ for $H$, with the exact $E_{\mathrm{kin}}$ and with $H_{\mathrm{rel}}(\gamma^{\ast})$ replaced by $H_{\mathrm{rel}}^{\mathrm{Gauss}}(\gamma^{\ast})$, \cref{eq:H_rel_main} evaluated with $S_0^{\gamma^{\ast}}$:
\begin{equation}
    H_{\mathrm{est}}=H_{\mathrm{rel}}^{\mathrm{Gauss}}(\gamma^{\ast})+\beta\gamma^{\ast}E_{\mathrm{kin}}+\ln\mathcal I[0;\gamma^{\ast}] .
\label{eq:SI_Hest}
\end{equation}
Its deviation, $H_{\mathrm{est}}-H=H_{\mathrm{rel}}^{\mathrm{Gauss}}(\gamma^{\ast})-H_{\mathrm{rel}}(\gamma^{\ast})$, is that of the closed form against $\hat P_0^{\gamma^{\ast}}$. \Cref{eq:SI_Hrel_gamma} at $\gamma=1$ then gives the relative entropy against the physical reference with the same absolute error.

This deviation has two sources. One is the Gaussian approximation. The other is the reference: relative to $\hat P_0^{\gamma^{\ast}}$, the Hamiltonian of $\hat P$ differs by $(1-\gamma^{\ast})\hat{\mathcal H}_0+\hat U$, and a density coupling cannot represent the kinetic part, so even the exact inference against $\hat P_0^{\gamma^{\ast}}$ does not return $\hat P$. Repeating the numerical inference of \cref{SI:control} against $\hat P_0^{\gamma^{\ast}}$ changes the entropy by at most $0.07\%$, less than $1\%$ of the closed-form deviation at $\beta t\geq2$.

\subsection{Numerical checks and convergence}
\label{SI:numerics}

The Matsubara sums of the closed form run over $|n|\leq150$. Their summand decays as $\omega_n^{-4}$ (\cref{SI:bridge}), and raising the cutoff to $|n|\leq1200$ changes the closed-form entropy by less than $4\times10^{-7}$ of $H$ at every benchmark point. Against the physical reference, $\gamma=1$, the summand tends to the constant $h(E_{\mathrm{kin}}/E^0_{\mathrm{kin}})\neq0$ [\cref{eq:SI_ratio_limit}], and we verified that the partial sums then grow linearly with the cutoff.

Sums of $S(q,\omega_n)$ itself converge slowly. Beyond a cutoff $n_c$, \cref{eq:tail_lattice} gives $S(q,\omega_n)\simeq C_q/\omega_n^2$ with $C_q=2(1-\cos q)|E_{\mathrm{kin}}|/N\hbar^2$, and
\begin{equation}
    \sum_{|n|>n_c}\frac{1}{\omega_n^2}=2\Big(\frac{\beta\hbar}{2\pi}\Big)^2\sum_{n>n_c}\frac{1}{n^2}=2\Big(\frac{\beta\hbar}{2\pi}\Big)^2\psi_1(n_c+1),
\label{eq:SI_tail_sum}
\end{equation}
with $\psi_1$ the trigamma function. Adding $C_q$ times \cref{eq:SI_tail_sum} to the sum over $|n|\leq n_c$ recovers $S(q)$ [\cref{eq:SI_static_from_Matsubara}] to six significant figures, whereas truncation alone leaves relative deviations of up to $8\times10^{-3}$. The Gaussian free energies of \cref{tab:SI_Fex} sum to $|n|=5000$ and add the tail in the same way, with $\ln(1+2N\lambda_qS_0)\simeq2N\lambda_qC^0_q/\omega_n^2$ and $C^0_q$ the coefficient of the free reference.

\begin{figure}[t]
\includegraphics[width=\linewidth]{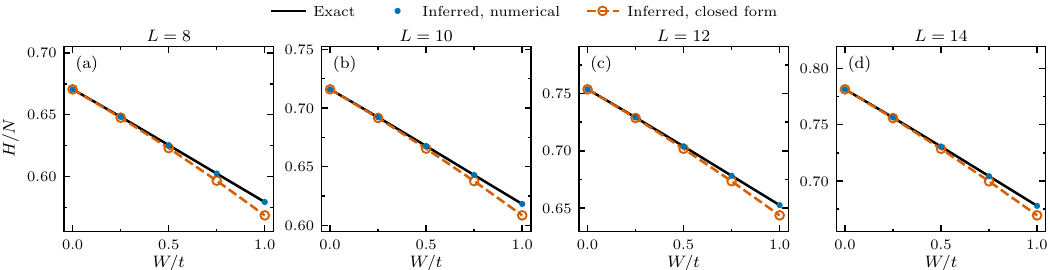}
\caption{Benchmark for spinless fermions on a ring, \cref{eq:tW_model}, at
half filling, $N=L/2$, and $\beta t=1$. Entropy per particle $H/N$
against interaction strength $W/t$ for (a) $L=8$, (b) $L=10$, (c) $L=12$, and
(d) $L=14$; the vertical scales differ between panels. The solid line is exact
diagonalization. Dots are numerical inference without the Gaussian approximation, which
reproduces the exact entropy at every point. Circles are the closed-form
estimate, \cref{eq:H_rel_main}, converted to $H$ with the exact kinetic energy.
Panel (c) repeats \cref{fig:benchmark}(b).}
\label{fig:SI_ring_size}
\end{figure}

The closed-form error depends only weakly on the ring size (\cref{fig:SI_ring_size}). At half filling, $N=L/2$, and $W=t$, the relative error in $H$ at $\beta t=1$ decreases from $1.9\%$ at $L=8$ to $1.2\%$ at $L=14$; at $\beta t=2$ it stays between $10\%$ and $13\%$.

At $L=12$, $N=6$, the lowest excitation energy rises from $t$ at $W=0$ to $1.3t$ at $W=t$, at least four times the thermal energy at $\beta t=4$, so $\beta t=4$ is effectively zero temperature for this ring.

\subsection{Exact free energies and the Gaussian approximation}
\label{SI:cancellation}

At the exact multipliers, $\lambda_q=W\cos q/L$ (\cref{SI:control}), the error of the Gaussian approximation can be measured directly: the normalization $\mathcal Z_\lambda$ of $\hat P_\lambda$ gives the exact excess free energy, and \cref{eq:SI_Fex_lambda} gives its Gaussian approximation, $\beta F_{\mathrm{ex}}'^{\,\mathrm{Gauss}}=\tfrac12\sum_{q\neq0,n}\ln(1+2N\lambda_{q}S_0)$, the random-phase ring-diagram sum (\cref{sec:rpa}). Here $F_{\mathrm{ex}}'=F_{\mathrm{ex}}-WN^2/L$ is the excess free energy of $\hat U-WN^2/L$ [\cref{eq:U_lattice}]; the constant does not affect $H$ or $H_{\mathrm{rel}}$. The Gaussian approximation deviates from the exact value by $1.5$--$19\%$ (\cref{tab:SI_Fex}).

By \cref{eq:SI_Hrel_instantaneous}, and in the Gaussian approximation by \cref{eq:SI_Hrel_lambda_static}, the relative entropy at these multipliers is the difference
\begin{equation}
    H_{\mathrm{rel}} = \beta N\sum_{q\neq0}\lambda_{q}S(q) - \beta F_{\mathrm{ex}}' .
\label{eq:SI_residue}
\end{equation}
The two terms cancel to $88$--$96\%$ (\cref{tab:SI_Fex}), so a small error in $\beta F_{\mathrm{ex}}'$ becomes a large error in $H_{\mathrm{rel}}$. The $1.5\%$ error at $W=0.25t$ gives $35\%$, and at $W=t$ the error reverses the sign of $H_{\mathrm{rel}}$. For the closed form of \cref{fig:benchmark}, $H$ and $H_{\mathrm{rel}}$ differ by exact terms [\cref{eq:SI_Hrel_gamma}], so they carry the same absolute error. At $\beta t=2$ and $W=t$ it is $10\%$ of $H$ and $46\%$ of $H_{\mathrm{rel}}$, against either reference.

\begin{table}[t]
\caption{Excess free energy $\beta F_{\mathrm{ex}}'=\beta(F_{\mathrm{ex}}-WN^2/L)$, exact and in the Gaussian approximation, at the exact multipliers $\lambda_{q}=W\cos q/L$, for \cref{eq:tW_model} at $L=12$, $N=6$, $\beta t=2$, with the free reference at hopping $t$. ``Cancellation'' is $1-|H_{\mathrm{rel}}|/|\beta N\sum_{q\neq0}\lambda_{q}S(q)|$ in \cref{eq:SI_residue}; for the inferred multipliers against the $\gamma^{\ast}$ reference used in \cref{fig:benchmark}, with $N\sum_{q\neq0,n}\lambda_{q,n}S(q,\omega_n)$ in place of $\beta N\sum_{q\neq0}\lambda_{q}S(q)$, it is $93.6\%$, $89.3\%$, $86.1\%$ and $83.5\%$.}
\label{tab:SI_Fex}
\begin{ruledtabular}
\begin{tabular}{ccccc}
$W/t$ & $\beta F_{\mathrm{ex}}'$ exact & $\beta F_{\mathrm{ex}}'$ Gaussian & error & cancellation \\
\hline
0.25 & $-0.5791$ & $-0.5879$ & $1.5\%$ & $95.8\%$ \\
0.5  & $-1.2079$ & $-1.2615$ & $4.4\%$ & $92.6\%$ \\
0.75 & $-1.8830$ & $-2.0613$ & $9.5\%$ & $90.0\%$ \\
1    & $-2.6018$ & $-3.0959$ & $19\%$  & $87.9\%$ \\
\end{tabular}
\end{ruledtabular}
\end{table}

\section{Uniform electron gas}
\label{sec:ueg}

We use Hartree atomic units, $\hbar=m_{\mathrm e}=e=1$, so that $\omega_n=2\pi n/\beta$ and $U_{\mathbf q}=4\pi/q^2$.

\subsection{State point and input data}
\label{SI:ueg_data}

The spin-unpolarized electron gas at density parameter $r_s$ has $\bar\rho=3/(4\pi r_s^3)$, Fermi wave number $k_{\mathrm F}=(9\pi/4)^{1/3}/r_s$ and Fermi energy $E_{\mathrm F}=k_{\mathrm F}^2/2$, and $\Theta=1$ fixes $\beta=1/E_{\mathrm F}$. \citet{chuna2025estimates} computed $f(\mathbf q,\tau)$, \cref{eq:SI_imaginary_time_density}, in a cubic cell of edge $L_{\mathrm c}=(4\pi N/3)^{1/3}r_s$ and volume $V=L_{\mathrm c}^3$, with periodic boundary conditions, from $M=1000$ independent simulations with a canonical worm-algorithm code. The deposited data~\cite{chuna_thomas_michael_2025_3923} give $f(\mathbf q,\tau)$ on $P=200$ imaginary-time slices, $\tau_j=j\beta/P$, together with its $M$ leave-one-out averages, each omitting one simulation. The statistical error of $f$, set by the fermion sign problem, is of order $10^{-2}$ at $r_s=2$, $10^{-3}$ at $r_s=5$ and $10^{-4}$ at $r_s=10$ and $20$~\cite{chuna2025estimates}. The data set also contains $r_s=50$, $100$ and $200$. These lie beyond the quantum Monte Carlo data to which the reference parametrization was fitted, and we do not use them.

The allowed wave vectors are $\mathbf q=(2\pi/L_{\mathrm c})(n_x,n_y,n_z)$ with integer $n_i$, and every sum over $\mathbf q\neq0$ on the cell runs over the lattice vectors of the $100$ shells with $n_x^2+n_y^2+n_z^2\leq118$, up to $q_{\mathrm{last}}=6.81k_{\mathrm F}$. Each deposited set has $100$ files, one per shell, but $15$ of them duplicate the preceding shell, so $15$ shells have no data. For these shells, the summand per vector is interpolated linearly in $q$ between the neighboring shells. These shells carry at most $1.3\%$ of $\beta\Delta F_{\mathrm{corr}}$ on the cell and $0.4\%$ of $f_{\mathrm c}$.

\subsection{Reference and Matsubara transform}
\label{SI:ueg_transform}

\Cref{eq:Fcorr} compares $S(\mathbf q,\omega_n)$ with $S_0(\mathbf q,\omega_n)$ mode by mode, so both are needed on the wave vectors of the cell. The reference is the ideal Fermi gas on the lattice of the cell: \cref{eq:S0_fermi} with $g_s=2$ and $\mathbf k$ on that lattice. Since $S_0(\mathbf q,\omega_n)$ is real (\cref{SI:density_correlations}), only the real part of each term of \cref{eq:S0_fermi} contributes,
\begin{equation}
    S_0(\mathbf q,\omega_n)=\frac{2}{N}\sum_{\mathbf k}\frac{(f_k-f_{k+q})(\epsilon_{k+q}-\epsilon_k)}{\omega_n^2+(\epsilon_{k+q}-\epsilon_k)^2} ,
\label{eq:SI_ueg_S0}
\end{equation}
with the degenerate pairs, $\epsilon_{k+q}=\epsilon_k$, replaced by their limit $\beta f_k(1-f_k)\,\delta_{n0}$.

The data give $f(\mathbf q,\tau)$ only on the slices $\tau_j$, so on each shell we interpolate it by a cubic spline $s(\tau)$ whose derivatives at $\tau=0$ and $\beta$ are fixed by the f-sum rule, $s'(0)=-q^2/2$ and $s'(\beta)=q^2/2$ (\cref{SI:high_frequency}), and integrate it exactly in \cref{eq:SI_Matsubara_transform}. Three integrations by parts, with $s(\beta)=s(0)$ and $s'''$ constant between knots, give, for $n\neq0$,
\begin{equation}
    S(\mathbf q,\omega_n)=\frac{q^2}{\omega_n^2}+\frac{1}{\omega_n^4}\sum_{k=0}^{P-1}J_k\cos(\omega_n\tau_k) ,
\label{eq:SI_ueg_spline}
\end{equation}
with $J_k$ the jump of $s'''$ at the knot $\tau_k$, the knots taken periodically, and $S(\mathbf q,0)=\int_0^\beta s(\tau)\,\mathrm d\tau$. The first term is the f-sum asymptote of \cref{eq:SI_fsum}, so the transform has the exact high-frequency limit. It reproduces $\beta^{-1}\sum_nS(\mathbf q,\omega_n)=f(\mathbf q,0)$ to $3\times10^{-6}$. At $r_s=20$ it agrees with the Matsubara density response that \citet{dornheim2024dynamic} obtained from their simulations of the same system~\cite{dornheim_tobias_2024_3088} to a median of $6\times10^{-4}$ over the shared shells and $|n|\leq29$.

\subsection{Static structure factor as input}
\label{SI:ueg_reconstruction}

To check whether the deviation of the closed form from the reference comes from the input, we also reconstruct $S(\mathbf q,\omega_n)$ from the simulated $S(\mathbf q)=f(\mathbf q,0)$ alone, as $S_\lambda$ of \cref{eq:SI_S_lambda}, and call it the static model. On each shell, \cref{eq:SI_stationarity_static} fixes $\lambda_{\mathbf q}$. We sum $|n|\leq300$ explicitly. Beyond it, every term of \cref{eq:SI_stationarity_static} equals its f-sum asymptote $q^2/\omega_n^2$ up to $O(\omega_n^{-4})$, independent of $\lambda_{\mathbf q}$ [\cref{eq:SI_fsum}], so the terms with $|n|>300$ add $\beta^{-1}q^2$ times the sum of \cref{eq:SI_tail_sum} with $n_c=300$. This remainder stays below $7\times10^{-3}$ on the fitted shells, far below $S(\mathbf q)$, and Brent's method finds the root.

At large $q$, $S_\lambda$ hardly depends on $\lambda_{\mathbf q}$, since $S_0/S_\lambda-1=2N\lambda_{\mathbf q}S_0$ [\cref{eq:SI_S_lambda}] vanishes, so the fit would turn the statistical error of $S(\mathbf q)$ into an error of $\lambda_{\mathbf q}$. We therefore fit only where the bare interaction changes the $n=0$ term by more than about $5\%$, $\bar\rho U_{\mathbf q}S_0(\mathbf q,0)>0.05$, that is, up to $q=2.5k_{\mathrm F}$ at $r_s=2$ and $4.5k_{\mathrm F}$ at $r_s=20$.

Beyond that shell, we take $\lambda_{\mathbf q}\propto U_{\mathbf q}$, with the ratio of the last fitted shell, and the model still reproduces $S(\mathbf q)$ to within $8\times10^{-3}$. With $U_{\mathbf q}\propto q^{-2}$ and $S_0\leq2/\epsilon_q$, $S_0/S_\lambda-1$ then falls off as $q^{-4}$. The summand of \cref{eq:Fcorr} is quadratic in it, and about $\beta\epsilon_q\propto q^2$ Matsubara terms contribute, so each mode contributes $O(q^{-6})$, and the sum over $\mathbf q$ converges.

The $f(\mathbf q,\beta/2)$ of the static model, from \cref{eq:SI_inverse_Matsubara}, deviates from the data on the shells with $q<4k_{\mathrm F}$ by up to $3\%$ at $r_s=5$ and $11\%$ at $r_s=20$, many times the statistical error. At $r_s=2$ the two agree within the errors of the data. Its closed-form $f_{\mathrm c}$ nevertheless differs from that of the data by no more than about $1\%$ (\cref{tab:SI_ueg}), so errors in the imaginary-time structure of the input cannot account for the deviation from the reference.

\subsection{Correlation free energy on the cell}
\label{SI:ueg_cell}

The summand of \cref{eq:Fcorr}, $\tfrac12(\ln y-y+1)$ with $y=S_0/S$, equals $-\tfrac14(y-1)^2+O[(y-1)^3]$, so statistical noise in $S$ lowers every term, by about a quarter of the variance of $S/S_0$, and the bias accumulates over the many lattice vectors at large $q$, where the signal is small. We remove it term by term with the jackknife estimate: with $z$ a term evaluated on all simulations and $\bar z$ its mean over the $M$ leave-one-out averages, the bias is $(M-1)(\bar z-z)$, an estimate that is exact for the $O(1/M)$ bias of a quadratic function of the mean. The leave-one-out averages also give the statistical error. The bias amounts to $13\%$ of $|f_{\mathrm c}|$ at $r_s=2$ and to at most $0.1\%$ at the other densities.

The Matsubara sum runs over $|n|\leq n_c=20$, because at $r_s\leq5$ the data at larger $|n|$ are dominated by noise, and their bias exceeds their signal. The omitted terms carry less than $10^{-5}$ of the sum in the static model of \cref{SI:ueg_reconstruction} and less than $0.2\%$ in the data at $r_s=10$ and $20$, where the noise is smallest. Varying $n_c$ between $10$ and $50$ changes $f_{\mathrm c}$ by at most $0.3\%$ at $r_s\geq5$, and at $r_s=2$ the values for $n_c=10$ and $20$ agree to $0.1\%$, well within the statistical error of $1.2\%$.

The static model has no noise, so its sum can be converged: it runs over $|n|\leq300$, plus an analytic remainder. At large $|n|$, $S=a/\omega_n^2+b/\omega_n^4+O(\omega_n^{-6})$, with $a=q^2$ for $S$ and $S_0$ alike [\cref{eq:SI_fsum}] and $b$ and $b_0$ the next coefficients of $S$ and $S_0$. Then $y-1=(b_0-b)/a\omega_n^2+O(\omega_n^{-4})$, the summand of \cref{eq:Fcorr} decays as $-(b_0-b)^2/4a^2\omega_n^4$, and the remainder beyond $|n|=300$ follows from $\sum_{|n|>300}\omega_n^{-4}=2(\beta/2\pi)^4\zeta(4,301)$, with $\zeta(s,x)=\sum_{k\geq0}(k+x)^{-s}$ the Hurwitz zeta function. We obtain $b$ and $b_0$ by a least-squares fit of $\omega_n^2S-a$ against $\omega_n^{-2}$ over the $80$ largest $|n|\leq300$. In the static model, $y-1=2N\lambda_{\mathbf q}S_0$ exactly [\cref{eq:SI_S_lambda}], so $(b_0-b)/a=2N\lambda_{\mathbf q}q^2$. The fitted values agree with this to a few parts in $10^5$ on the inner shells and to $3\times10^{-3}$ on the outermost shell. Lowering the cutoff to $150$ changes the static-model $f_{\mathrm c}$ by less than $10^{-7}$ of its value.

\subsection{Finite-size correction}
\label{SI:ueg_finite_size}

Because the cell has no modes with $q<2\pi/L_{\mathrm c}$, samples the remaining range coarsely, and stops at the last reported shell, we correct the cell value to the thermodynamic limit. In this limit, the sum over $\mathbf q$ in \cref{eq:Fcorr} becomes $(V/2\pi^2)\int\mathrm dq\,q^2$, since the summand depends only on $q=|\mathbf q|$. The data do not give $S(\mathbf q,\omega_n)$ between and outside their shells. We therefore evaluate the integral and the cell sum in the random-phase approximation and add their difference to the cell value, as in finite-size corrections of quantum Monte Carlo data~\cite{chiesa2006finite,dornheim2016ab}. This approximation becomes exact at long wavelengths and sets $2V\lambda_{\mathbf q}=U_{\mathbf q}$. By \cref{eq:SI_S_lambda}, $S_0/S=1+\bar\rho U_{\mathbf q}S_0$, so the summand of \cref{eq:Fcorr} becomes that of \cref{eq:F_corr_rpa}, and the correction is
\begin{equation}
    \beta\,\delta F=\bigg[\frac{V}{2\pi^2}\int_0^{q_{\max}}\mathrm dq\,q^2-\sum_{\mathbf q\neq0}^{\text{cell}}\bigg]\,\frac12\sum_n\Big[\ln\big(1+\bar\rho U_{\mathbf q}S_0\big)-\bar\rho U_{\mathbf q}S_0\Big],
\label{eq:SI_ueg_fs}
\end{equation}
with $q_{\max}=12k_{\mathrm F}$, beyond the last shell of the data (\cref{SI:ueg_data}). Both terms use the continuum Lindhard function, at the chemical potential of the ideal gas of density $\bar\rho$. With $\sum_{\mathbf k}\to V\int\mathrm d\mathbf k/(2\pi)^3$ in \cref{eq:SI_ueg_S0}, the identity $\sum_{\mathbf k}(f_k-f_{k+q})\,g(\epsilon_{k+q}-\epsilon_k)=2\sum_{\mathbf k}f_k\,g(\epsilon_{k+q}-\epsilon_k)$ for odd $g$, and $\epsilon_{k+q}-\epsilon_k=q^2/2+kq\cos\theta$, the angular integral gives
\begin{equation}
    S_0(q,\omega_n)=\frac{1}{2\pi^2\bar\rho\,q}\int_0^\infty\mathrm dk\,k\,f_k\,\ln\frac{\omega_n^2+(q^2/2+kq)^2}{\omega_n^2+(q^2/2-kq)^2} .
\label{eq:SI_ueg_S0_continuum}
\end{equation}
The $k$ integral uses composite Gauss--Legendre quadrature with $16$ nodes per panel, on $36$ equal panels plus panels that shrink geometrically toward $k=q/2$, where the static ($n=0$) integrand of \cref{eq:SI_ueg_S0_continuum} has a logarithmic singularity. The $q$ integral uses $240$ Gauss--Legendre nodes. The frequency sums run over $|n|\leq200$, with the remainder treated as in \cref{SI:ueg_cell} and the fit over the $40$ largest $|n|$. The closed-form estimate is then $f_{\mathrm c}=(\Delta F_{\mathrm{corr}}^{\text{cell}}+\delta F)/N$. On the shells, the continuum Lindhard function agrees with the lattice one of the cell value to $4\times10^{-5}$, which changes $f_{\mathrm c}$ by less than $10^{-5}$ of its value.

The correction (\cref{tab:SI_ueg}) has three parts: the modes below $2\pi/L_{\mathrm c}=0.63k_{\mathrm F}$; the difference between the integral and the lattice sum over the sampled range; and the random-phase contribution between $q_{\mathrm{last}}$ and $q_{\max}$. The first part alone is $82\%$ of the closed-form $f_{\mathrm c}$ at $r_s=2$ and $65\%$ at $r_s=20$, the second reduces $|f_{\mathrm c}|$ by $14\%$ and $17\%$, and the third is below $0.5\%$.

Since the correction is a large part of $f_{\mathrm c}$, we test how much it depends on the model by replacing the random-phase approximation with the static model. By \cref{eq:U_lambda_lfc}, its multipliers define a static local field correction, $2V\lambda_{\mathbf q}=U_{\mathbf q}[1-G(\mathbf q)]$, so working with $\lambda_{\mathbf q}$ or with $G$ is equivalent. The static model changes the first two parts in opposite directions. Below the first shell the data do not fix $\lambda_{\mathbf q}$, and we use the compressibility sum rule~\cite{ichimaru1982strongly,giuliani2005quantum} instead. By this rule, $2V\lambda_{\mathbf q}-U_{\mathbf q}=-U_{\mathbf q}G$ tends to $\partial^2(\bar\rho f_{\mathrm{xc}})/\partial\bar\rho^2$ as $q\to0$, with the reference $f_{\mathrm{xc}}$ differentiated at fixed temperature. We take $2V\lambda_{\mathbf q}=U_{\mathbf q}+\partial^2(\bar\rho f_{\mathrm{xc}})/\partial\bar\rho^2-bq^2$, with $b$ fixed by the static-model value at the first shell. The constant does not vanish, so even as $q\to0$ the zero-frequency term of each mode differs from its random-phase value. At the first shell the sum rule alone gives $2V\lambda_{\mathbf q}/U_{\mathbf q}=0.874$ and $0.860$ for $r_s=2$ and $20$, close to the static-model values $0.874$ and $0.868$. These multipliers reduce $|f_{\mathrm c}|$ by $3.6\%$ at $r_s=2$ and by $3.0\%$ at $r_s=20$. In the second part, the static model, with $\lambda_{\mathbf q}/U_{\mathbf q}$ interpolated linearly between shells, increases $|f_{\mathrm c}|$ by $2.5\%$ and $2.1\%$. With the static model in both parts, $|f_{\mathrm c}|$ decreases by $1.1\%$ at $r_s=2$ and by $0.9\%$ at $r_s=20$, away from the reference, in line with \citet{dornheim2016ab}, who found a similar insensitivity for the potential energy.

\subsection{Reference values and error budget}
\label{SI:ueg_reference}

The exchange free energy per electron is the exact first-order result at finite temperature~\cite{horovitz1974exchange},
\begin{equation}
    f_{\mathrm x}=-\frac{2r_s^3}{3\pi^2\beta^2}\int_{-\infty}^{\beta\mu_0}\mathrm d\eta\,I_{-1/2}(\eta)^2,\qquad I_\nu(\eta)=\int_0^\infty\mathrm dx\,\frac{x^\nu}{e^{x-\eta}+1},
\label{eq:SI_ueg_fx}
\end{equation}
with $\mu_0$ the chemical potential of the ideal gas, fixed by $\bar\rho=(\sqrt2/\pi^2)\beta^{-3/2}I_{1/2}(\beta\mu_0)$. The exchange-correlation free energy $f_{\mathrm{xc}}$ is taken from the parametrization of \citet{karasiev2014accurate}, with the corrected parameters of Ref.~\cite{karasiev2018nonempirical} (Table S2 of its Supplemental Material).

We check this parametrization against that of Groth \emph{et al.}~\cite{groth2017ab}, because the finite-temperature data behind the corrected parameters end at $r_s=10$. Groth \emph{et al.} fitted data for $0.1\leq r_s\leq20$ and $0.5\leq\Theta\leq8$ and give an accuracy of about $0.3\%$ in $f_{\mathrm{xc}}$. Since $f_{\mathrm x}$ is exact, this is about $0.5\%$ of $|f_{\mathrm c}|$ at our state points. Both parametrizations were fitted to largely the same quantum Monte Carlo data for $\Theta\geq0.5$~\cite{karasiev2018nonempirical}, and at $\Theta=1$ and $2\leq r_s\leq20$ they agree to within $0.17\%$ in $f_{\mathrm{xc}}$ and $0.23\%$ in $f_{\mathrm c}$. With the parametrization of Groth \emph{et al.} as the reference, the deviations in \cref{tab:SI_ueg} change by at most $0.2\%$ of $|f_{\mathrm c}|$. \Cref{tab:SI_ueg} collects the values plotted in \cref{fig:fc}.

To find the source of the deviation from the reference, $8.0\%$ to $13.3\%$ of $|f_{\mathrm c}|$, we compare it with the other sources of error. The statistical error is $1.2\%$ of $|f_{\mathrm c}|$ at $r_s=2$ and below $0.1\%$ at the other densities, and varying the Matsubara cutoff between $n_c=10$ and $50$ changes $f_{\mathrm c}$ by at most $0.3\%$ at $r_s\geq5$ (\cref{SI:ueg_cell}). The shells without data carry at most $0.4\%$ of $f_{\mathrm c}$ (\cref{SI:ueg_data}). Replacing the random-phase approximation in the finite-size correction by the static model changes $f_{\mathrm c}$ by about $1\%$ (\cref{SI:ueg_finite_size}), and replacing the simulated imaginary-time structure by that model changes $f_{\mathrm c}$ by no more than about $1\%$ (\cref{SI:ueg_reconstruction}). The shells beyond $3k_{\mathrm F}$, where the growth of \cref{eq:Fcorr} with the momentum cutoff dominates, contribute $1.2\%$ of $|f_{\mathrm c}|$ at $r_s=20$ and $0.4\%$ at $r_s=10$, and this growth is itself a property of the Gaussian approximation (\cref{SI:ueg_cutoff}). The reference is known to about $0.5\%$ of $|f_{\mathrm c}|$. The deviation exceeds each of these effects by a factor of six or more, so it comes mainly from the Gaussian approximation.

\begin{table}[t]
\caption{Correlation free energy per electron of the spin-unpolarized uniform electron gas at $\Theta=1$, in Hartree. Columns: the exchange free energy, \cref{eq:SI_ueg_fx}; the closed form, \cref{eq:Fcorr}, on the $N=34$ cell from the simulated $f(\mathbf q,\tau)$; the finite-size correction, \cref{eq:SI_ueg_fs}; their sum, the closed-form $f_{\mathrm c}$ plotted in \cref{fig:fc}; the reference $f_{\mathrm c}=f_{\mathrm{xc}}-f_{\mathrm x}$; the relative deviation of the closed form from the reference, in $f_{\mathrm c}$ and in $f_{\mathrm{xc}}=f_{\mathrm x}+f_{\mathrm c}$; the share of the finite-size correction in the closed-form $f_{\mathrm c}$; and the deviation in $f_{\mathrm c}$ with the static model of \cref{SI:ueg_reconstruction} in place of the data. Statistical errors of the last digits are given in parentheses where they reach them.}
\label{tab:SI_ueg}
\begin{ruledtabular}
\begin{tabular}{cccccccccc}
$r_s$ & $f_{\mathrm x}$ & cell & correction & $f_{\mathrm c}$ & reference $f_{\mathrm c}$ & dev.\ in $f_{\mathrm c}$ & dev.\ in $f_{\mathrm{xc}}$ & correction share & static model \\
\hline
2  & $-0.086925$ & $-0.0415(17)$ & $-0.088405$ & $-0.1299(17)$ & $-0.141212$ & $8.0(12)\%$ & $4.9\%$ & $68\%$ & $8.5\%$ \\
5  & $-0.034770$ & $-0.02653(5)$ & $-0.039004$ & $-0.06554(5)$ & $-0.074175$ & $11.6\%$ & $7.9\%$ & $60\%$ & $11.6\%$ \\
10 & $-0.017385$ & $-0.01758(2)$ & $-0.020385$ & $-0.03797(2)$ & $-0.043711$ & $13.1\%$ & $9.4\%$ & $54\%$ & $13.5\%$ \\
20 & $-0.008693$ & $-0.011001(3)$ & $-0.010507$ & $-0.021508(3)$ & $-0.024802$ & $13.3\%$ & $9.8\%$ & $49\%$ & $14.2\%$ \\
\end{tabular}
\end{ruledtabular}
\end{table}

\subsection{Large wave numbers and the momentum cutoff}
\label{SI:ueg_cutoff}

The sum over $\mathbf q$ in \cref{eq:Fcorr} grows linearly with the momentum cutoff $q_{\max}$. At large $q$, the interaction-induced change in kinetic energy makes $S/S_0-1$ proportional to $1/\epsilon_q$. Each mode then contributes a term proportional to $1/\epsilon_q\propto q^{-2}$, and a shell of width $\mathrm dq$ holds a number of modes proportional to $q^2\,\mathrm dq$, so every shell adds the same amount.

\emph{Exact correlation functions at large $q$.} By \cref{eq:SI_density_mode,eq:SI_imaginary_time_density},
\begin{equation}
    f(\mathbf q,\tau)=\frac{1}{N}\sum_{i,j=1}^{N}\big\langle e^{\tau\hat{\mathcal H}}e^{-i\mathbf q\cdot\hat{\mathbf r}_i}e^{-\tau\hat{\mathcal H}}e^{i\mathbf q\cdot\hat{\mathbf r}_j}\big\rangle .
\end{equation}
At $\tau=0$ the terms with $i\neq j$ add up to $S(\mathbf q)-1$, which vanishes as $q^{-4}$ at large $q$ because the pair distribution has a cusp at contact~\cite{giuliani2005quantum}; we drop these terms, which affects only higher orders in $1/q$. For $i=j$, the momentum shift $e^{-i\mathbf q\cdot\hat{\mathbf r}_j}\hat{\mathbf p}_je^{i\mathbf q\cdot\hat{\mathbf r}_j}=\hat{\mathbf p}_j+\mathbf q$, together with the fact that the potential commutes with $e^{i\mathbf q\cdot\hat{\mathbf r}_j}$, gives $e^{-i\mathbf q\cdot\hat{\mathbf r}_j}\hat{\mathcal H}e^{i\mathbf q\cdot\hat{\mathbf r}_j}=\hat{\mathcal H}+\hat A_j$ with $\hat A_j=\mathbf q\cdot\hat{\mathbf p}_j+\epsilon_q$, so this term is the average of $e^{\tau\hat{\mathcal H}}e^{-\tau(\hat{\mathcal H}+\hat A_j)}=\exp\big(-\tau\hat A_j-\tfrac{1}{2}\tau^2[\hat{\mathcal H},\hat A_j]+O(\tau^3)\big)$, by the Baker--Campbell--Hausdorff formula. The commutator $[\hat{\mathcal H},\hat A_j]=[\hat U,\mathbf q\cdot\hat{\mathbf p}_j]=i\mathbf q\cdot\nabla_j\hat U$ is of order $q$, against $q^2$ for $\hat A_j$. Over the times that matter, $\tau\lesssim1/\epsilon_q$, it changes the result by a relative amount of order $\tau^2q\sim q^{-3}$. To the order needed, the term is therefore $e^{-\epsilon_q\tau}\langle e^{-\tau\mathbf q\cdot\hat{\mathbf p}_j}\rangle$, an average over the momentum distribution of the interacting state. Odd powers of $\mathbf q\cdot\hat{\mathbf p}_j$ average to zero by isotropy, and $\langle(\mathbf q\cdot\hat{\mathbf p}_j)^2\rangle=q^2\langle\hat{\mathbf p}_j^2\rangle/3=4\epsilon_qE_{\mathrm{kin}}/3N$, so that
\begin{equation}
    f(\mathbf q,\tau)=e^{-\epsilon_q\tau}\Big[1+\frac{2\epsilon_qE_{\mathrm{kin}}}{3N}\,\tau^2+O\big(q^4\tau^4\big)+O\big(q^{-3}\big)\Big],\qquad \tau\lesssim1/\epsilon_q .
\label{eq:SI_ueg_f_largeq}
\end{equation}

Since $f(\mathbf q,\beta-\tau)=f(\mathbf q,\tau)$ [\cref{eq:SI_f_symmetric}] and $f$ decays on the scale $1/\epsilon_q\ll\beta$, \cref{eq:SI_Matsubara_transform} becomes $S(\mathbf q,\omega_n)=2\,\mathrm{Re}\int_0^\infty\mathrm d\tau\,e^{i\omega_n\tau}f(\mathbf q,\tau)$, up to terms exponentially small in $\beta\epsilon_q$. With $\int_0^\infty\mathrm d\tau\,\tau^ke^{-(\epsilon_q-i\omega_n)\tau}=k!/(\epsilon_q-i\omega_n)^{k+1}$, $\mathrm{Re}\,(\epsilon_q-i\omega_n)^{-1}=\epsilon_q/(\epsilon_q^2+\omega_n^2)$ and $\mathrm{Re}\,(\epsilon_q-i\omega_n)^{-3}=\epsilon_q(\epsilon_q^2-3\omega_n^2)/(\epsilon_q^2+\omega_n^2)^3$,
\begin{equation}
    S(\mathbf q,\omega_n)=\frac{2}{\epsilon_q}\,\frac{1}{1+X^2}\bigg[1+\frac{4E_{\mathrm{kin}}}{3N\epsilon_q}\,\frac{1-3X^2}{(1+X^2)^2}+O\big(q^{-3}\big)\bigg],\qquad X=\frac{\omega_n}{\epsilon_q} .
\end{equation}
The ideal gas obeys the same expansion with $E_{\mathrm{kin}}^{0}$, so that
\begin{equation}
    \frac{S(\mathbf q,\omega_n)}{S_0(\mathbf q,\omega_n)}-1=\frac{4t_{\mathrm{xc}}}{3\epsilon_q}\,\frac{1-3X^2}{(1+X^2)^2}+O\big(q^{-3}\big),\qquad t_{\mathrm{xc}}\equiv\frac{E_{\mathrm{kin}}-E_{\mathrm{kin}}^{0}}{N},
\label{eq:SI_ueg_largeq}
\end{equation}
where $t_{\mathrm{xc}}$ is the exchange-correlation kinetic energy per electron. At fixed $n$, $X\to0$ and $S/S_0-1\to4t_{\mathrm{xc}}/3\epsilon_q$. Written as $S_0/S-1=\bar\rho U_{\mathbf q}[1-G(\mathbf q,i\omega_n)]S_0$ [\cref{eq:U_lambda_lfc}], with $S_0\to2/\epsilon_q$, this is a local field correction $G\simeq2t_{\mathrm{xc}}q^2/3\omega_{\mathrm p}^2$, with $\omega_{\mathrm p}^2=4\pi\bar\rho$ the squared plasma frequency, the short-wavelength limit derived in Refs.~\cite{hou2022exchange,dornheim2025short}.

\emph{Consequence for the sum over $\mathbf q$.} The summand of \cref{eq:Fcorr} is $-\tfrac14(y-1)^2$ to leading order (\cref{SI:ueg_cell}), with $y-1=S_0/S-1=-(S/S_0-1)+O(q^{-4})$. For $\beta\epsilon_q\gg1$ the Matsubara sum becomes an integral, $\sum_n\to(\beta\epsilon_q/2\pi)\int\mathrm dX$, and with $\int\mathrm dX\,(1-3X^2)^2/(1+X^2)^4=\pi/2$ each mode contributes
\begin{equation}
    \frac12\sum_{n}\Big[\ln\frac{S_0}{S}-\frac{S_0}{S}+1\Big]=-\frac{\beta t_{\mathrm{xc}}^2}{9\epsilon_q}+O\big(q^{-3}\big) .
\label{eq:SI_ueg_mode}
\end{equation}
A shell of width $\mathrm dq$ holds $Vq^2\mathrm dq/2\pi^2$ modes, so the modes with $q<q_{\max}$ give a sum that grows linearly with $q_{\max}$,
\begin{equation}
    \frac{\mathrm d(\beta\Delta F_{\mathrm{corr}})}{\mathrm dq_{\max}}\;\xrightarrow{q_{\max}\to\infty}\;-\frac{\beta Vt_{\mathrm{xc}}^2}{9\pi^2},
    \qquad
    \frac{\mathrm d(\Delta F_{\mathrm{corr}}/N)}{\mathrm d(q_{\max}/k_{\mathrm F})}\;\xrightarrow{q_{\max}\to\infty}\;-\frac{t_{\mathrm{xc}}^2}{6E_{\mathrm F}},
\label{eq:SI_ueg_slope}
\end{equation}
where the second form uses $V/N=3\pi^2/k_{\mathrm F}^3$. For the exact inference, $\Delta F_{\mathrm{corr}}=-\beta^{-1}D(\hat P_0\|\hat P)$ [\cref{eq:SI_Fcorr_general}] is finite, so the linear growth comes from the Gaussian approximation at large $q$. We therefore stop the sum at the last simulated shell, $q_{\mathrm{last}}=6.81k_{\mathrm F}$ (\cref{SI:ueg_data}).

\emph{Size at the state points.} At the exact multipliers, $F_{\text{ex}}/N=f_{\mathrm{xc}}$ (main text), so by \cref{eq:kinetic_shift_mass} $t_{\mathrm{xc}}=-m\,\partial f_{\mathrm{xc}}/\partial m$ at fixed $T$ and $V$. Restoring units, $f_{\mathrm{xc}}=(me^4/\hbar^2)\,\phi(r_s,\Theta)$, and both $r_s=\bar a\,me^2/\hbar^2$ and $\Theta=k_{\mathrm B}T/E_{\mathrm F}$ are proportional to $m$, so $m\partial_m f_{\mathrm{xc}}=f_{\mathrm{xc}}+r_s\partial_{r_s}f_{\mathrm{xc}}+\Theta\partial_\Theta f_{\mathrm{xc}}$. Hence
\begin{equation}
    t_{\mathrm{xc}}=-f_{\mathrm{xc}}-r_s\frac{\partial f_{\mathrm{xc}}}{\partial r_s}\bigg|_{\Theta}-\Theta\frac{\partial f_{\mathrm{xc}}}{\partial\Theta}\bigg|_{r_s} .
\label{eq:SI_ueg_txc}
\end{equation}
At $\Theta=1$, with the parametrization of \cref{SI:ueg_reference}, $t_{\mathrm{xc}}$ rises from $-0.017E_{\mathrm F}$ at $r_s=2$ to $0.336E_{\mathrm F}$ at $r_s=20$ and changes sign at $r_s=3.3$, as noted in Ref.~\cite{dornheim2025short}. By \cref{eq:SI_ueg_slope}, the exact $S(\mathbf q,\omega_n)$ would lower $\Delta F_{\mathrm{corr}}/N$ by $0.016\%$ of $|f_{\mathrm c}|$ per unit increase of $q_{\max}/k_{\mathrm F}$ at $r_s=2$ and by $0.35\%$ at $r_s=20$. Over the outer $2k_{\mathrm F}$ of the cell, the cell sum from the simulation input falls by $0.34\%$ of $|f_{\mathrm c}|$ per unit $q_{\max}/k_{\mathrm F}$ at $r_s=20$ and by $0.09\%$ at $r_s=10$, against the predicted $0.35\%$ and $0.14\%$. At $r_s=2$ and $5$ the predicted growth lies below the statistical noise.

\emph{The static model.} With $S_0=(2/\epsilon_q)(1+X^2)^{-1}$ and $\int\mathrm dX\,(1-3X^2)/(1+X^2)^3=0$, the $q^{-2}$ term of \cref{eq:SI_ueg_largeq} adds nothing to $\beta^{-1}\sum_nS(\mathbf q,\omega_n)$, so $S(\mathbf q)$ does not depend on $t_{\mathrm{xc}}$ at order $q^{-2}$. The static model, fixed by $S(\mathbf q)$, has no such term, and its sum over $\mathbf q$ converges (\cref{SI:ueg_reconstruction}).

\end{document}